\documentclass[twocolumn]{aastex631}

\shorttitle{GC System Sizes}
\shortauthors{Lim et al.}
\graphicspath{{./}{figures/}}

\begin{document}

\title{The Spatial Distribution of Globular Cluster Systems in Early Type Galaxies: Estimation Procedure and Catalog of Properties for Globular Cluster Systems Observed with Deep Imaging Surveys}

\correspondingauthor{Sungsoon Lim}
\email{sslim00@gmail.com}
 
\author[0000-0002-5049-4390]{Sungsoon Lim}
\affiliation{Department of Astronomy, Yonsei University, 50 Yonsei-ro, Seodaemun-gu, Seoul 03722, Republic of Korea}

\author[0000-0002-2073-2781]{Eric W. Peng}
\affiliation{NSF’s NOIRLab, 950 N. Cherry Avenue, Tucson, AZ 85719, USA}

\author[0000-0003-1184-8114]{Patrick C{\^o}t{\'e}}
\affiliation{Herzberg Astronomy and Astrophysics Research Centre, National Research Council of Canada, Victoria, BC V9E 2E7, Canada}

\author[0000-0002-8224-1128]{Laura Ferrarese}
\affiliation{Herzberg Astronomy and Astrophysics Research Centre, National Research Council of Canada, Victoria, BC V9E 2E7, Canada}

\author[0000-0002-0363-4266]{Joel C. Roediger}
\affiliation{Herzberg Astronomy and Astrophysics Research Centre, National Research Council of Canada, Victoria, BC V9E 2E7, Canada}

\author[0000-0002-4718-3428]{Chengze Liu}
\affiliation{Department of Astronomy, School of Physics and Astronomy, and Shanghai Key Laboratory for Particle Physics and Cosmology, Shanghai Jiao Tong University, Shanghai 200240, China}

\author[0000-0002-1685-4284]{Chelsea Spengler}
\affiliation{Institute of Astrophysics, Pontificia Universidad Cat\'olica de Chile, Av. Vicu\~{n}a Mackenna 4860, 7820436 Macul, Santiago, Chile}

\author[0000-0002-2814-3578]{Elisabeth Sola}
\affiliation{Institute of Astronomy, Madingley Rd, Cambridge, CB3 0HA, UK}

\author[0000-0003-3343-6284]{Pierre-Alain Duc}
\affiliation{Universit{\'e} de Strasbourg, CNRS, Observatoire astronomique de Strasbourg (ObAS), UMR 7550, 67000 Strasbourg, France}

\author[0000-0002-3790-720X]{Laura V. Sales}
\affiliation{Department of Physics and Astronomy, University of California, Riverside, 900 University Avenue, Riverside, CA 92521, USA)}

\author[0000-0002-5213-3548]{John P. Blakeslee}
\affiliation{NSF’s NOIRLab, 950 N. Cherry Avenue, Tucson, AZ 85719, USA}

\author[0000-0002-3263-8645]{Jean-Charles Cuillandre}
\affiliation{AIM Paris Saclay, CNRS/INSU, CEA/Irfu, Université Paris Diderot, Orme des Merisiers, F-91191 Gif-sur-Yvette Cedex, France}

\author[0000-0001-9427-3373]{Patrick R. Durrell}
\affiliation{Department of Physics and Astronomy, Youngstown State University, One University Plaza, Youngstown, OH 44555, USA}

\author[0000-0002-6155-7166]{Eric Emsellem}
\affiliation{European Southern Observatory, Karl-Schwarzschild Stra{\ss}e 2, D-85748 Garching bei M\"{u}nchen, Germany}

\author[0000-0001-8221-8406]{Stephen D. J. Gwyn}
\affiliation{Herzberg Astronomy and Astrophysics Research Centre, National Research Council of Canada, Victoria, BC V9E 2E7, Canada}

\author[0000-0002-7214-8296]{Ariane Lan{\c c}on}
\affiliation{Universit{\'e} de Strasbourg, CNRS, Observatoire astronomique de Strasbourg (ObAS), UMR 7550, 67000 Strasbourg, France}

\author[0000-0002-1442-2947]{Francine R. Marleau}
\affiliation{Institut f{\"u}r Astro- und Teilchenphysik, Universit{\"a}t Innsbruck, Technikerstra{\ss}e 25/8, Innsbruck, A-6020, Austria}

\author[0000-0002-7089-8616]{J. Christopher Mihos}
\affiliation{Department of Astronomy, Case Western Reserve University, Cleveland, OH 44106, USA}

\author[0000-0003-4552-9808]{Oliver M\"uller}
\affiliation{Institute of Physics, Laboratory of Astrophysics, École Polytechnique Fédérale de Lausanne (EPFL), 1290 Sauverny, Switzerland}

\author[0000-0003-0350-7061]{Thomas H. Puzia}
\affiliation{Institute of Astrophysics, Pontificia Universidad Cat\'olica de Chile, Av. Vicu\~{n}a Mackenna 4860, 7820436 Macul, Santiago, Chile}

\author[0000-0003-4945-0056]{Rub{\'e}n S{\'a}nchez-Janssen}
\affiliation{STFC UK Astronomy Technology Centre, Royal Observatory, Blackford Hill, Edinburgh, EH9 3HJ, UK}



\begin{abstract}

We present an analysis of the spatial distribution of globular cluster (GC) systems of 118 nearby early-type galaxies in the Next Generation Virgo Cluster Survey (NGVS) and Mass Assembly of early-Type GaLAxies with their fine Structures (MATLAS) survey programs, which both used MegaCam on the Canada-France-Hawaii Telescope.  
We describe the procedure used to select GC candidates and fit the spatial distributions of GCs to a two-dimensional S\'ersic function, which provides effective radii (half number radii) and S\'ersic indices, and estimate background contamination by adding a constant term to the S'ersic function.  
In cases where a neighboring galaxy affects the estimation of the GC spatial distribution in the target galaxy, we fit two 2D S\'ersic functions, simultaneously. 
We also investigate the color distributions of GCs in our sample by using Gaussian Mixture Modeling. 
For GC systems with bimodal color distributions, we divide the GCs into blue and red subgroups and fit their respective spatial distributions with S\'ersic functions. 
Finally, we measure the total number of GCs based on our fitted S\'ersic function, and calculate the GC specific frequency.

\end{abstract}



\section{Introduction} \label{sec:intro}

Globular clusters (GCs) have long been used as probes of galaxy formation and evolution. 
While studies of GCs belonging to individual galaxies have provided many insights into galaxy evolution, it is the systematic study of GCs in large surveys 
(e.g., \citealp{2004ApJS..153..223C,2007ApJ...670.1074M,2007ApJS..169..213J,2009MNRAS.392..879G,2012AJ....144..154R,2014ApJ...796...52B,2015ApJ...799..159Z,2021MNRAS.503.2406A})
that allow us to explore the general relationships between GC systems and their host galaxies. 
For example, it is well known that GC color distributions and total numbers (or total masses) are closely related to properties of their host galaxies (e.g., \citealp{1997AJ....114..482B,2006ApJ...639...95P,2008ApJ...681..197P,2009MNRAS.392L...1S,2017ApJ...836...67H}).
Although GC colors and numbers have been used to study galaxy formation and evolution with empirical relations (e.g., \citealp{2010Sci...328..334L,2011ApJ...730...23P,2011A&A...525A..19C,2018ApJ...862...82L,2019MNRAS.484.4865P,2020ApJ...899...69L,2023ApJ...953..154H,2023ApJS..265....9H}) much less is known about the {\it spatial distribution} of GC systems as this requires
deep, wide-field imaging from galaxy cores to their peripheries.

Early efforts focused on estimating GC ``spatial extents", taken to be the point where the GC number density profile merges into the background (e.g., \citealp{2007AJ....134.1403R,2012AJ....144..164H,2014ApJ...796...62H}). While this concept of `GC extent' can be helpful in understanding how far GCs extend from the galaxy center, a homogeneous comparison of results from different surveys can be problematic.

For this reason, functional modeling of GC spatial distributions is generally needed to avoid dependence on surveys. 
There have been wide-field imaging surveys for GCs in individual, or handfuls, of galaxies that estimated effective radii for GC systems (e.g., \citealp{2012MNRAS.420...37B,2013MNRAS.436.1172U,2014ApJ...797..128H,2014MNRAS.437..273K,2016MNRAS.458..105K}), but sample sizes have been too small to fully explore the link between galaxy properties and GC system size.
Several recent studies have reported effective radii for GC systems (and other galaxy properties) but there are often discrepancies among the results; moreover, samples have been still limited to $20-30$ galaxies --- usually massive systems \citep{2018MNRAS.477.3869H,2017MNRAS.472L.104F,2019MNRAS.488.4504C}.
A few studies estimating GC spatial distributions with functional models for a large sample of galaxies exist (e.g., \citealp{2015ApJ...799..159Z,2017ApJ...849....6A}), but they have not focused on the spatial distributions of GC systems. 
Therefore, a systematic study of GC spatial distributions for a larger, more representative sample of galaxies is needed to better understand GC spatial distributions.

In this study, we examine the spatial distribution of GCs belonging to 118 early-type galaxies based on imaging from two large nearby galaxy surveys --- the Next Generation Virgo Cluster Survey (NGVS, PI: Ferrarese, L., \citealp{2012ApJS..200....4F}) and Mass Assembly of early-Type GaLAxies with their fine Structures (MATLAS, PI: Duc, P.-A., \citealp{2020arXiv200713874D}). 
Additionally, our analysis also uses Hubble Space Telescope imaging from the ACS Virgo Cluster Survey (PI: Cote, P.). 
We note that the scientific analysis and interpretation of the results are published in \citet{2024ApJ...966..168L} so the focus of this paper is on the data products. 
In \S2, we describe our data and methods, including photometry and fitting for analytic functional form. 
In \S3, the results of individual galaxies are presented and discussed. 
We summarize our results in \S4.

\section{Data and Methodology} 

\subsection{Target selection}
The data used in this study were obtained from two large optical imaging surveys: NGVS and MATLAS.
While imaging for numerous galaxies is available from these surveys, we limited our targets to ensure reliable GC studies, mainly focusing on early-type galaxies (ETGs).
First, we targeted nearby ($\leq 25$ Mpc) MATLAS galaxies (including galaxies inside the NGVS footprint) that were observed in at least three filters.
Second, we also targeted ACSVCS galaxies inside the NGVS footprint to include low-mass (early-type dwarf) galaxies in our sample. 
Among the galaxies that satisfy the above categories, several systems were too close to neighboring, giant galaxies with their own rich GC systems, making it almost impossible to detect and study GCs in these galaxies; such objects were excluded for our analysis. 
Ultimately, we targeted 118 galaxies in this study.
Table \ref{tbl:galaxies} lists our targets and their properties.

\startlongtable
\begin{deluxetable*}{crrccccccc}
\tablenum{1}
\tablecaption{List of sample galaxies\label{tbl:galaxies}. For MATLAS galaxies, magnitudes and colors (columns 4,5) are from Sola et al. (2022); galaxy stellar masses (column 6) are taken from \citet{2013MNRAS.432.1862C}; 
and effective radii (column 7) are results of \citet{2011MNRAS.413..813C}. For NGVS galaxies, magnitudes, colors, and effective radii (columns 4,5,7) are taken from \citet{2020ApJ...890..128F}; and galaxy stellar masses (column 6) are from Roediger et al. (2024, in preparation). The distances (column 8) are based on HST surface brightness fluctuation measurements for ACSVCS galaxies \citep{2007ApJ...655..144M,2009ApJ...694..556B}, while distances
for MATLAS and NGVS (not in ACSVCS) galaxies are taken
from \citet{2011MNRAS.413..813C} 
} 
\tablewidth{0pt}
\tablehead{
\colhead{Name} & \colhead{RA (J2000)} &\colhead{Dec (J2000)} & \colhead{$M_g$} & \colhead{$(g'-i')_0$} & \colhead{ $M_{*}$} &$R_{e,*}$ &\colhead{Distance} & \colhead{Survey} & \colhead{Other name} \\
\colhead{} & \colhead{[degrees] } & \colhead{[degrees] } & \colhead{[mag]} & \colhead{[mag]} & \colhead{[$M_{\odot}$]} & \colhead{[$\arcsec$]} & \colhead{[Mpc]} & \colhead{} & \colhead{}
}
\decimalcolnumbers
\startdata
NGC0524 & $21.198778$ & $9.538793$ & $-20.82$ & $1.00$ & $1.2 \times 10^{11}$ & $43.7$ & $23.3$ & MATLAS & ... \\
NGC0821 & $32.088123$ & $10.994870$ & $-20.62$ & $0.81$ & $7.9 \times 10^{10}$ & $39.8$ & $23.4$ & MATLAS & ... \\
NGC0936 & $36.906090$ & $-1.156280$ & $-21.16$ & $1.01$ & $9.8 \times 10^{10}$ & $52.5$ & $22.4$ & MATLAS & ... \\
NGC1023 & $40.100052$ & $39.063251$ & $-20.20$ & $0.96$ & $5.1 \times 10^{10}$ & $47.9$ & $11.1$ & MATLAS & ... \\
NGC2592 & $126.783669$ & $25.970339$ & $-19.07$ & $1.03$ & $3.7 \times 10^{10}$ & $12.3$ & $25.0$ & MATLAS & ... \\
NGC2685 & $133.894791$ & $58.734409$ & $-19.18$ & $0.80$ & $1.2 \times 10^{10}$ & $25.7$ & $16.7$ & MATLAS & ... \\
NGC2768 & $137.906265$ & $60.037209$ & $-20.91$ & $1.03$ & $2.7 \times 10^{11}$ & $63.1$ & $21.8$ & MATLAS & ... \\
NGC2778 & $138.101639$ & $35.027424$ & $-18.73$ & $0.92$ & $2.4 \times 10^{10}$ & $15.8$ & $22.3$ & MATLAS & ... \\
NGC2950 & $145.646317$ & $58.851219$ & $-19.43$ & $0.93$ & $2.1 \times 10^{10}$ & $15.5$ & $14.5$ & MATLAS & ... \\
NGC3098 & $150.569458$ & $24.711092$ & $-19.23$ & $0.91$ & $1.7 \times 10^{10}$ & $13.2$ & $23.0$ & MATLAS & ... \\
NGC3245 & $156.826523$ & $28.507435$ & $-20.09$ & $0.92$ & $4.5 \times 10^{10}$ & $25.1$ & $20.3$ & MATLAS & ... \\
NGC3379 & $161.956665$ & $12.581630$ & $-20.07$ & $1.04$ & $5.0 \times 10^{10}$ & $39.8$ & $10.3$ & MATLAS & M105 \\
NGC3384 & $162.070404$ & $12.629300$ & $-19.89$ & $0.93$ & $2.4 \times 10^{10}$ & $32.4$ & $11.3$ & MATLAS & ... \\
NGC3457 & $163.702591$ & $17.621157$ & $-18.56$ & $0.88$ & $4.0 \times 10^{9}$ & $13.5$ & $20.1$ & MATLAS & ... \\
NGC3489 & $165.077454$ & $13.901258$ & $-19.60$ & $0.85$ & $8.6 \times 10^{9}$ & $22.4$ & $11.7$ & MATLAS & ... \\
NGC3599 & $168.862305$ & $18.110369$ & $-19.02$ & $0.88$ & $7.2 \times 10^{9}$ & $23.4$ & $19.8$ & MATLAS & ... \\
NGC3607 & $169.227737$ & $18.051809$ & $-21.19$ & $0.96$ & $1.6 \times 10^{11}$ & $38.9$ & $22.2$ & MATLAS & ... \\
NGC3608 & $169.245697$ & $18.148531$ & $-19.99$ & $0.96$ & $6.4 \times 10^{10}$ & $29.5$ & $19.8$ & MATLAS & ... \\
NGC3630 & $170.070786$ & $2.964170$ & $-19.55$ & $0.95$ & $2.5 \times 10^{10}$ & $12.6$ & $25.0$ & MATLAS & ... \\
NGC3945 & $178.307190$ & $60.675560$ & $-20.72$ & $1.03$ & $6.9 \times 10^{10}$ & $28.2$ & $23.2$ & MATLAS & ... \\
IC3032 & $182.782333$ & $14.274944$ & $-15.96$ & $0.77$ & $6.3 \times 10^{8}$ & $9.0$ & $15.0$ & NGVS,ACSVCS & VCC33 \\
IC3065 & $183.802417$ & $14.433083$ & $-17.11$ & $0.83$ & $2.0 \times 10^{9}$ & $9.2$ & $16.5$ & NGVS,ACSVCS & VCC140 \\
VCC200 & $184.140333$ & $13.031417$ & $-16.70$ & $0.86$ & $5.9 \times 10^{8}$ & $12.9$ & $18.3$ & NGVS,ACSVCS & ... \\
IC3101 & $184.331833$ & $11.943389$ & $-15.92$ & $0.86$ & $4.4 \times 10^{8}$ & $9.4$ & $17.9$ & NGVS,ACSVCS & VCC230 \\
NGC4262 & $184.877426$ & $14.877717$ & $-19.00$ & $1.04$ & $1.9 \times 10^{10}$ & $8.5$ & $15.5$ & NGVS,ACSVCS & VCC355 \\
NGC4267 & $184.938675$ & $12.798356$ & $-19.66$ & $1.06$ & $3.9 \times 10^{10}$ & $28.9$ & $15.8$ & NGVS,ACSVCS & VCC369 \\
NGC4278 & $185.028320$ & $29.280619$ & $-20.16$ & $1.02$ & $7.5 \times 10^{10}$ & $31.6$ & $15.6$ & MATLAS & ... \\
NGC4283 & $185.086609$ & $29.310898$ & $-18.21$ & $1.02$ & $8.3 \times 10^{9}$ & $12.3$ & $15.3$ & MATLAS & ... \\
UGC7436 & $185.581458$ & $14.760722$ & $-17.00$ & $0.84$ & $2.1 \times 10^{9}$ & $18.2$ & $15.8$ & NGVS,ACSVCS & VCC543 \\
VCC571 & $185.671417$ & $7.950306$ & $-17.02$ & $0.80$ & $6.3 \times 10^{8}$ & $10.6$ & $23.8$ & NGVS,ACSVCS & ... \\
NGC4318 & $185.680458$ & $8.198250$ & $-18.05$ & $0.95$ & $3.3 \times 10^{9}$ & $5.8$ & $22.0$ & NGVS,ACSVCS & VCC575 \\
NGC4339 & $185.895599$ & $6.081713$ & $-19.18$ & $1.03$ & $1.9 \times 10^{10}$ & $24.8$ & $16.0$ & NGVS & VCC648 \\
NGC4340 & $185.897141$ & $16.722195$ & $-19.70$ & $1.01$ & $2.3 \times 10^{10}$ & $29.0$ & $18.4$ & NGVS,ACSVCS & VCC654 \\
NGC4342 & $185.912598$ & $7.053936$ & $-18.52$ & $1.15$ & $1.2 \times 10^{10}$ & $4.5$ & $16.5$ & NGVS & VCC657 \\
NGC4350 & $185.990891$ & $16.693356$ & $-19.69$ & $1.08$ & $3.4 \times 10^{10}$ & $15.1$ & $15.4$ & NGVS,ACSVCS & VCC685 \\
NGC4352 & $186.020833$ & $11.218333$ & $-18.42$ & $0.96$ & $6.2 \times 10^{9}$ & $15.6$ & $18.5$ & NGVS,ACSVCS & VCC698 \\
NGC4365 & $186.117615$ & $7.317520$ & $-22.02$ & $1.04$ & $1.3 \times 10^{11}$ & $75.4$ & $23.1$ & NGVS,ACSVCS & VCC731 \\
NGC4371 & $186.230957$ & $11.704288$ & $-19.99$ & $1.05$ & $3.9 \times 10^{10}$ & $28.5$ & $16.9$ & NGVS,ACSVCS & VCC759 \\
NGC4374 & $186.265747$ & $12.886960$ & $-22.05$ & $1.06$ & $2.2 \times 10^{11}$ & $90.4$ & $18.5$ & NGVS,ACSVCS & M84,VCC763 \\
NGC4377 & $186.301285$ & $14.762218$ & $-19.04$ & $0.99$ & $1.5 \times 10^{10}$ & $10.7$ & $17.7$ & NGVS,ACSVCS & VCC778 \\
NGC4379 & $186.311386$ & $15.607498$ & $-18.89$ & $1.01$ & $1.6 \times 10^{10}$ & $13.7$ & $15.9$ & NGVS,ACSVCS & VCC784 \\
NGC4387 & $186.423813$ & $12.810359$ & $-18.72$ & $1.02$ & $1.1 \times 10^{10}$ & $10.8$ & $18.0$ & NGVS,ACSVCS & VCC828 \\
IC3328 & $186.490875$ & $10.053556$ & $-17.21$ & $0.86$ & $2.1 \times 10^{9}$ & $17.4$ & $16.9$ & NGVS,ACSVCS & VCC856 \\
NGC4406 & $186.549225$ & $12.945970$ & $-22.24$ & $1.00$ & $2.6 \times 10^{11}$ & $135.8$ & $17.9$ & NGVS,ACSVCS & M86,VCC881 \\
NGC4417 & $186.710938$ & $9.584117$ & $-19.50$ & $1.01$ & $2.5 \times 10^{10}$ & $15.3$ & $16.0$ & NGVS,ACSVCS & VCC944 \\
NGC4425 & $186.805664$ & $12.734803$ & $-18.77$ & $1.02$ & $1.3 \times 10^{10}$ & $16.9$ & $16.5$ & NGVS & VCC984 \\
NGC4429 & $186.860657$ & $11.107540$ & $-20.72$ & $1.08$ & $8.8 \times 10^{10}$ & $42.8$ & $16.5$ & NGVS & VCC1003 \\
NGC4434 & $186.902832$ & $8.154311$ & $-19.32$ & $0.98$ & $1.1 \times 10^{10}$ & $12.1$ & $22.5$ & NGVS,ACSVCS & VCC1025 \\
NGC4435 & $186.918762$ & $13.079021$ & $-20.15$ & $1.03$ & $4.3 \times 10^{10}$ & $25.6$ & $16.7$ & NGVS,ACSVCS & VCC1030 \\
NGC4442 & $187.016220$ & $9.803620$ & $-20.04$ & $1.07$ & $5.2 \times 10^{10}$ & $17.8$ & $15.3$ & NGVS,ACSVCS & VCC1062 \\
IC3383 & $187.051208$ & $10.297500$ & $-16.41$ & $0.87$ & $9.3 \times 10^{8}$ & $18.6$ & $16.2$ & NGVS,ACSVCS & VCC1075 \\
IC3381 & $187.062083$ & $11.790000$ & $-18.06$ & $0.88$ & $4.2 \times 10^{9}$ & $40.4$ & $16.7$ & NGVS,ACSVCS & VCC1087 \\
NGC4452 & $187.180417$ & $11.755000$ & $-18.43$ & $0.95$ & $9.0 \times 10^{9}$ & $15.1$ & $15.6$ & NGVS,ACSVCS & VCC1125 \\
NGC4458 & $187.239716$ & $13.241916$ & $-18.76$ & $0.96$ & $1.0 \times 10^{10}$ & $21.9$ & $16.3$ & NGVS,ACSVCS & VCC1146 \\
NGC4459 & $187.250107$ & $13.978580$ & $-20.47$ & $1.03$ & $7.2 \times 10^{10}$ & $41.0$ & $16.0$ & NGVS,ACSVCS & VCC1154 \\
NGC4461 & $187.262543$ & $13.183857$ & $-19.59$ & $1.03$ & $2.9 \times 10^{10}$ & $18.6$ & $16.5$ & NGVS & VCC1158 \\
VCC1185 & $187.347625$ & $12.450667$ & $-16.05$ & $0.87$ & $7.6 \times 10^{8}$ & $19.4$ & $16.9$ & NGVS,ACSVCS & ... \\
NGC4472 & $187.444992$ & $8.000410$ & $-22.65$ & $1.02$ & $3.7 \times 10^{11}$ & $225.6$ & $16.7$ & NGVS,ACSVCS & M49,VCC1226 \\
NGC4473 & $187.453659$ & $13.429320$ & $-20.42$ & $1.02$ & $5.7 \times 10^{10}$ & $32.9$ & $15.2$ & NGVS,ACSVCS & VCC1231 \\
NGC4474 & $187.473099$ & $14.068673$ & $-19.03$ & $0.94$ & $1.4 \times 10^{10}$ & $20.0$ & $15.5$ & NGVS,ACSVCS & VCC1242 \\
NGC4476 & $187.496170$ & $12.348669$ & $-18.82$ & $0.91$ & $7.3 \times 10^{9}$ & $18.1$ & $17.7$ & NGVS,ACSVCS & VCC1250 \\
NGC4477 & $187.509048$ & $13.636443$ & $-20.25$ & $1.06$ & $5.5 \times 10^{10}$ & $33.7$ & $16.5$ & NGVS & VCC1253 \\
NGC4482 & $187.543292$ & $10.779472$ & $-18.10$ & $0.85$ & $4.1 \times 10^{9}$ & $20.1$ & $18.2$ & NGVS,ACSVCS & VCC1261 \\
NGC4478 & $187.572662$ & $12.328578$ & $-19.40$ & $1.01$ & $1.4 \times 10^{10}$ & $12.3$ & $17.1$ & NGVS,ACSVCS & VCC1279 \\
NGC4479 & $187.576667$ & $13.578028$ & $-18.24$ & $1.01$ & $7.2 \times 10^{9}$ & $17.6$ & $17.4$ & NGVS,ACSVCS & VCC1283 \\
NGC4483 & $187.669250$ & $9.015665$ & $-18.46$ & $0.98$ & $7.5 \times 10^{9}$ & $12.6$ & $16.7$ & NGVS,ACSVCS & VCC1303 \\
NGC4486 & $187.705933$ & $12.391100$ & $-22.23$ & $1.01$ & $2.9 \times 10^{11}$ & $105.0$ & $16.7$ & NGVS,ACSVCS & M87,VCC1316 \\
NGC4489 & $187.717667$ & $16.758696$ & $-18.39$ & $0.92$ & $8.7 \times 10^{9}$ & $17.8$ & $15.4$ & NGVS,ACSVCS & VCC1321 \\
IC3461 & $188.011208$ & $11.890222$ & $-16.36$ & $0.89$ & $1.1 \times 10^{9}$ & $11.6$ & $16.8$ & NGVS,ACSVCS & VCC1407 \\
NGC4503 & $188.025803$ & $11.176434$ & $-19.58$ & $1.06$ & $3.3 \times 10^{10}$ & $21.7$ & $16.5$ & NGVS & VCC1412 \\
IC3468 & $188.059208$ & $10.251389$ & $-17.79$ & $0.86$ & $3.5 \times 10^{9}$ & $29.2$ & $15.4$ & NGVS,ACSVCS & VCC1422 \\
IC3470 & $188.097375$ & $11.262833$ & $-16.90$ & $0.97$ & $2.3 \times 10^{9}$ & $10.1$ & $16.0$ & NGVS,ACSVCS & VCC1431 \\
IC798 & $188.139125$ & $15.415333$ & $-16.55$ & $0.88$ & $9.8 \times 10^{8}$ & $8.6$ & $16.1$ & NGVS,ACSVCS & VCC1440 \\
NGC4515 & $188.270625$ & $16.265528$ & $-18.16$ & $0.91$ & $5.5 \times 10^{9}$ & $9.7$ & $16.7$ & NGVS,ACSVCS & VCC1475 \\
VCC1512 & $188.394000$ & $11.261889$ & $-15.95$ & $0.80$ & $3.0 \times 10^{8}$ & $12.9$ & $18.3$ & NGVS,ACSVCS & ... \\
IC3501 & $188.465083$ & $13.322583$ & $-16.81$ & $0.92$ & $1.7 \times 10^{9}$ & $10.1$ & $16.3$ & NGVS,ACSVCS & VCC1528 \\
NGC4528 & $188.525269$ & $11.321266$ & $-18.59$ & $1.02$ & $1.2 \times 10^{10}$ & $8.9$ & $15.7$ & NGVS,ACSVCS & VCC1537 \\
VCC1539 & $188.528208$ & $12.741694$ & $-15.48$ & $1.00$ & $4.1 \times 10^{8}$ & $17.1$ & $17.0$ & NGVS,ACSVCS & ... \\
IC3509 & $188.548083$ & $12.048861$ & $-16.32$ & $1.06$ & $1.1 \times 10^{9}$ & $9.9$ & $16.8$ & NGVS,ACSVCS & VCC1545 \\
NGC4550 & $188.877548$ & $12.220955$ & $-18.89$ & $1.03$ & $1.6 \times 10^{10}$ & $11.4$ & $15.3$ & NGVS,ACSVCS & VCC1619 \\
NGC4551 & $188.908249$ & $12.264010$ & $-18.70$ & $1.04$ & $1.4 \times 10^{10}$ & $13.8$ & $16.2$ & NGVS,ACSVCS & VCC1630 \\
NGC4552 & $188.916183$ & $12.556040$ & $-21.07$ & $1.03$ & $9.6 \times 10^{10}$ & $59.2$ & $16.0$ & NGVS,ACSVCS & M89,VCC1632 \\
VCC1661 & $189.103375$ & $10.384611$ & $-15.40$ & $0.90$ & $5.3 \times 10^{8}$ & $18.7$ & $15.8$ & NGVS,ACSVCS & ... \\
NGC4564 & $189.112473$ & $11.439320$ & $-19.55$ & $1.01$ & $2.3 \times 10^{10}$ & $16.4$ & $15.9$ & NGVS,ACSVCS & VCC1664 \\
NGC4570 & $189.222504$ & $7.246663$ & $-19.95$ & $1.04$ & $3.5 \times 10^{10}$ & $14.5$ & $17.1$ & NGVS,ACSVCS & VCC1692 \\
NGC4578 & $189.377274$ & $9.555121$ & $-19.28$ & $0.96$ & $1.9 \times 10^{10}$ & $25.6$ & $16.4$ & NGVS,ACSVCS & VCC1720 \\
NGC4596 & $189.983063$ & $10.176031$ & $-20.46$ & $1.00$ & $5.0 \times 10^{10}$ & $42.3$ & $16.5$ & NGVS & VCC1813 \\
VCC1826 & $190.046833$ & $9.896083$ & $-15.61$ & $0.87$ & $5.3 \times 10^{8}$ & $6.6$ & $16.3$ & NGVS,ACSVCS & ... \\
VCC1833 & $190.081875$ & $15.935333$ & $-16.76$ & $0.83$ & $1.5 \times 10^{9}$ & $7.8$ & $16.3$ & NGVS,ACSVCS & ... \\
IC3647 & $190.221250$ & $10.476111$ & $-16.91$ & $0.60$ & $2.0 \times 10^{8}$ & $37.7$ & $16.2$ & NGVS,ACSVCS & VCC1857 \\
IC3652 & $190.243917$ & $11.184556$ & $-17.19$ & $0.91$ & $2.4 \times 10^{9}$ & $20.3$ & $16.1$ & NGVS,ACSVCS & VCC1861 \\
NGC4608 & $190.305374$ & $10.155793$ & $-19.71$ & $0.96$ & $3.0 \times 10^{10}$ & $26.6$ & $16.5$ & NGVS & VCC1869 \\
IC3653 & $190.315500$ & $11.387083$ & $-16.98$ & $0.97$ & $2.6 \times 10^{9}$ & $7.0$ & $15.5$ & NGVS,ACSVCS & VCC1871 \\
NGC4612 & $190.386490$ & $7.314782$ & $-19.50$ & $0.91$ & $1.6 \times 10^{10}$ & $25.1$ & $16.5$ & NGVS,ACSVCS & VCC1883 \\
VCC1886 & $190.414208$ & $12.247889$ & $-15.89$ & $0.69$ & $2.1 \times 10^{8}$ & $14.1$ & $15.7$ & NGVS,ACSVCS & ... \\
UGC7854 & $190.466667$ & $9.402861$ & $-16.24$ & $0.83$ & $8.4 \times 10^{8}$ & $10.3$ & $15.9$ & NGVS,ACSVCS & VCC1895 \\
NGC4621 & $190.509674$ & $11.646930$ & $-21.02$ & $1.02$ & $9.7 \times 10^{10}$ & $69.1$ & $14.9$ & NGVS,ACSVCS & M59,VCC1903 \\
NGC4638 & $190.697632$ & $11.442459$ & $-19.62$ & $0.96$ & $2.0 \times 10^{10}$ & $12.5$ & $17.5$ & NGVS,ACSVCS & VCC1938 \\
NGC4649 & $190.916702$ & $11.552610$ & $-21.99$ & $1.06$ & $2.5 \times 10^{11}$ & $76.0$ & $16.5$ & NGVS,ACSVCS & M60,VCC1978 \\
VCC1993 & $191.050083$ & $12.941694$ & $-15.87$ & $0.85$ & $6.2 \times 10^{8}$ & $11.1$ & $16.6$ & NGVS,ACSVCS & ... \\
NGC4660 & $191.133209$ & $11.190533$ & $-19.32$ & $1.02$ & $2.5 \times 10^{10}$ & $10.8$ & $15.0$ & NGVS,ACSVCS & VCC2000 \\
IC3735 & $191.335083$ & $13.692500$ & $-16.98$ & $0.83$ & $1.5 \times 10^{9}$ & $16.6$ & $17.2$ & NGVS,ACSVCS & VCC2019 \\
IC3773 & $191.813833$ & $10.203611$ & $-17.07$ & $0.85$ & $2.7 \times 10^{9}$ & $14.1$ & $13.5$ & NGVS,ACSVCS & VCC2048 \\
IC3779 & $191.836208$ & $12.166306$ & $-16.03$ & $0.84$ & $8.3 \times 10^{8}$ & $11.3$ & $15.8$ & NGVS,ACSVCS & VCC2050 \\
NGC4694 & $192.062881$ & $10.983624$ & $-19.39$ & $0.72$ & $8.1 \times 10^{9}$ & $25.6$ & $16.5$ & NGVS & VCC2066 \\
NGC4710 & $192.412323$ & $15.165490$ & $-19.94$ & $0.99$ & $4.5 \times 10^{10}$ & $25.2$ & $16.5$ & NGVS & ... \\
NGC4733 & $192.778259$ & $10.912103$ & $-18.63$ & $0.91$ & $1.1 \times 10^{10}$ & $26.3$ & $14.5$ & NGVS & VCC2087 \\
NGC4754 & $193.073181$ & $11.313660$ & $-20.14$ & $1.04$ & $3.7 \times 10^{10}$ & $26.1$ & $16.1$ & NGVS,ACSVCS & VCC2092 \\
NGC4762 & $193.233536$ & $11.230800$ & $-20.93$ & $1.02$ & $7.1 \times 10^{10}$ & $31.9$ & $22.6$ & NGVS,ACSVCS & VCC2095 \\
NGC5839 & $226.364471$ & $1.634633$ & $-18.83$ & $1.00$ & $2.0 \times 10^{10}$ & $16.6$ & $22.0$ & MATLAS & ... \\
NGC5846 & $226.621887$ & $1.605637$ & $-21.05$ & $1.06$ & $2.5 \times 10^{11}$ & $58.9$ & $24.2$ & MATLAS & ... \\
NGC5866 & $226.623169$ & $55.763309$ & $-20.59$ & $0.95$ & $7.8 \times 10^{10}$ & $36.3$ & $14.9$ & MATLAS & ... \\
PGC058114 & $246.517838$ & $2.906550$ & $-18.29$ & $0.78$ & $6.3 \times 10^{9}$ & $9.3$ & $23.8$ & MATLAS & ... \\
NGC6548 & $271.496826$ & $18.587217$ & $-19.59$ & $1.00$ & $4.6 \times 10^{10}$ & $22.4$ & $22.4$ & MATLAS & ... \\
NGC7280 & $336.614899$ & $16.148266$ & $-19.26$ & $0.92$ & $1.4 \times 10^{10}$ & $21.4$ & $23.7$ & MATLAS & ... \\
NGC7332 & $339.352173$ & $23.798351$ & $-20.28$ & $0.86$ & $2.4 \times 10^{10}$ & $17.4$ & $22.4$ & MATLAS & ... \\
NGC7457 & $345.249725$ & $30.144892$ & $-18.93$ & $0.89$ & $7.4 \times 10^{9}$ & $36.3$ & $12.9$ & MATLAS & ... \\
NGC7454 & $345.277130$ & $16.388371$ & $-19.10$ & $0.96$ & $2.9 \times 10^{10}$ & $25.7$ & $23.2$ & MATLAS & ... \\
\enddata
\tablecomments{Please be noted that there is a slight difference between the absolute g'-magnitudes of the NGVS samples and those presented in Fig. 1 of \citet{2024ApJ...966..168L}. In \citet{2024ApJ...966..168L}, we employed model-fitted magnitudes; however, in this study, we have chosen for curve-of-growth magnitudes, which we consider to be a more comprehensive magnitude for estimating the fluxes of galaxies.}
\end{deluxetable*}

\subsection{The Next Generation Virgo Cluster Survey}

The NGVS is a deep, multi-band imaging survey of the Virgo cluster carried out with MegaCam \citep{2003SPIE.4841...72B} on the Canada-France-Hawaii Telescope (CFHT) from 2008 to 2013. 
The survey covers an area of 104 deg$^2$ (with 117 pointings) within the virial radii of both the Virgo A and Virgo B subclusters. 
Full survey details, including observing strategy and data processing, are described in \citet{2012ApJS..200....4F}. 
Additional details on the data reduction and analysis procedures are also available in \citet{2020ApJ...890..128F}.

\subsection{The Mass Assembly of early-Type GaLAxies with their fine Structures}
The MATLAS survey is a second deep imaging survey using MegaCam on CFHT. The targets of the MATLAS survey are galaxies from the ATLAS$^{\rm 3D}$ sample \citep{2011MNRAS.413..813C}. The sample contains 260 nearby (within 42 Mpc) bright ($M_K <  -21.5$) early-type galaxies. Full survey details are available in \citet{2015MNRAS.446..120D,2020arXiv200713874D,2020MNRAS.498.2138B}, and the data reduction process is described in \citet{2008PASP..120..212G}.

\subsection{Photometry}
We used the NGVS aperture photometry catalog for the NGVS samples. 
The catalog details are fully described in \citet{2015ApJ...812L...2L}, so we present here only a brief description of the catalog.
Source Extractor \citep{1996A&AS..117..393B} was run on the processed images to obtain aperture magnitudes of sources with dual-image mode.
We used the g'-band images as detection images and adopted circular apertures with a series of diameters between 2 and 16 pixels to measure the source fluxes, which were then corrected to 16-pixel-diameter aperture magnitudes. Instrumental magnitudes were then calibrated to standard AB magnitudes through a comparison to SDSS PSF magnitudes after conversion to MegaCam filter magnitudes. 

The central regions of some galaxies have high surface brightness, making it difficult to estimate the background and detect sources with general photometry programs such as Source Extractor. 
Since all NGVS galaxies have been modeled with customized two-dimensional isophote (ISO) fitting models \citep{2020ApJ...890..128F}, we subtracted diffuse light from galaxies using ISO fit models to enhance source detection and background estimation. 
The model subtraction is performed with a cutout images having a $10\arcmin \times 10\arcmin$ field of view.
We then ran Source Extractor on these galaxy-subtracted images in the same way as for the NGVS source catalog.
The magnitudes measured on the model-subtracted images are then matched with the original NGVS catalog by comparing magnitudes of sources in the outer $2\arcmin$ width area of model-subtracted images. 
We replaced the NGVS aperture photometry catalogs of the central $8\arcmin\times8\arcmin$ regions of our target galaxies with the photometric catalogs on the model-subtracted images.

We also generated aperture photometry catalogs for MATLAS galaxies using the same methodology as the NGVS aperture photometry catalog. However, the detection images and galaxy model subtraction for MATLAS galaxies differ slightly from those for NGVS galaxies. For the detection image, we chose the best seeing filter image for MATLAS. We used ring median filtered galaxy models instead of two-dimensional ISO fit models for the model subtraction. 
The ring median filtering method can produce diffuse images by setting the inner and outer ring sizes. We set the radii of the inner and outer rings to $15$ and $20$ pixels, respectively. We subtracted these ring-median-filtered model galaxies from the MATLAS cutout images with a field of view of $10\arcmin \times 10\arcmin$, similar to the NGVS data. Additionally, we replaced the MATLAS aperture photometry catalogs for the central $8\arcmin\times8\arcmin$ regions of our target galaxies with photometric catalogs based on the model-subtracted images.

\subsection{Globular cluster selection}

\begin{figure}
\epsscale{1.1}
\plottwo{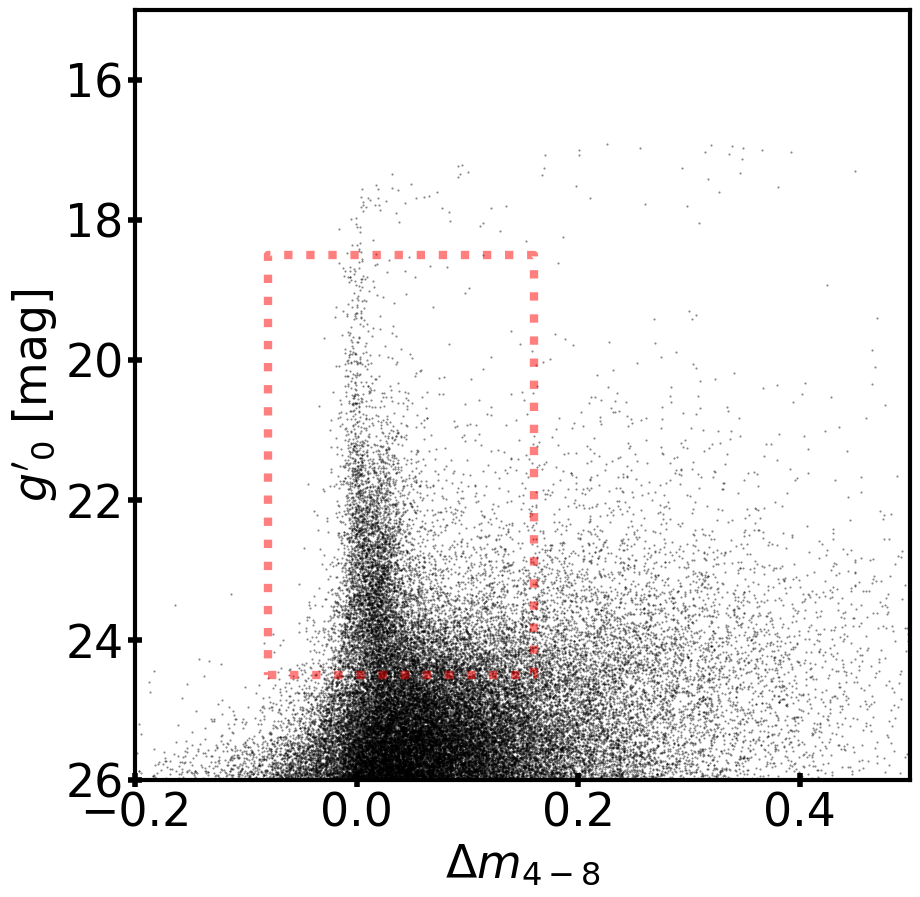}{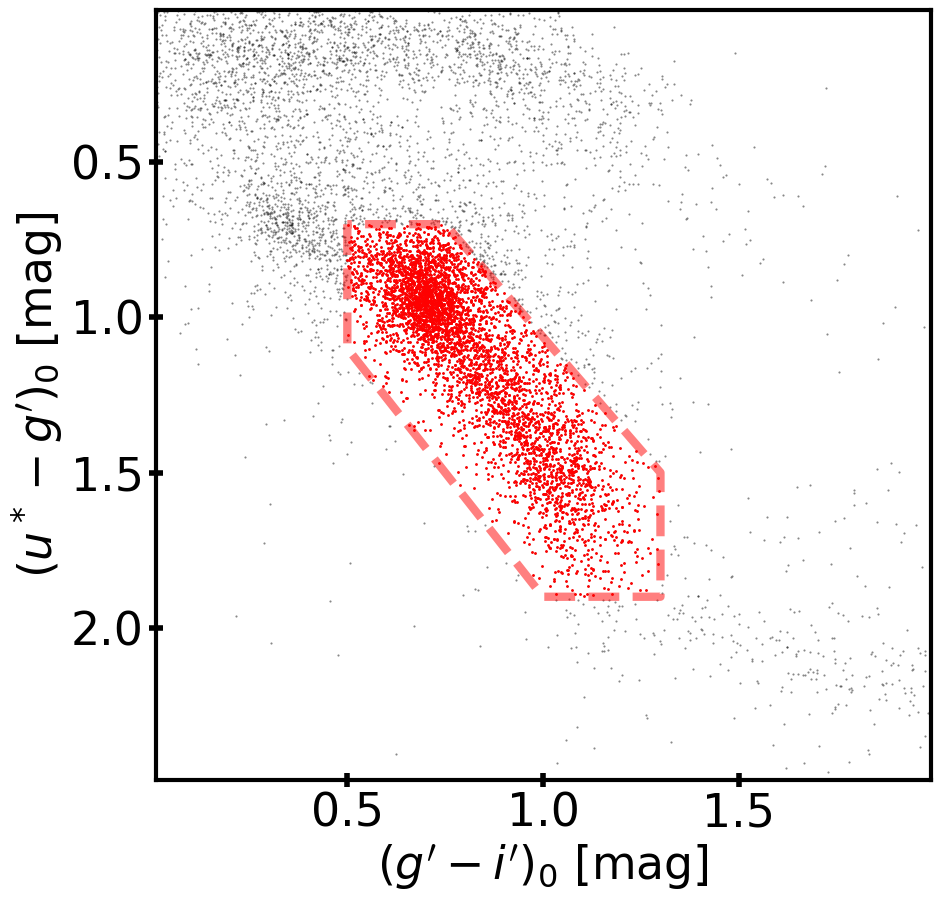}
\caption{(Left) Inverse concentration index, $\Delta m_{4-8}$, versus g-band magnitude for sources in the NGC4472 region. The red dotted box shows the point-like source region used for this galaxy. (Right) ($u^*-g'$)-($g'-i'$) color-color diagram of point-like sources in the NGC4472 region. The red dashed polygon shows the globular cluster (GC) selection region used in this study, with red sources showing GC candidates.
\label{gcselection}}
\end{figure}

\begin{figure}
\epsscale{1.1}
\plottwo{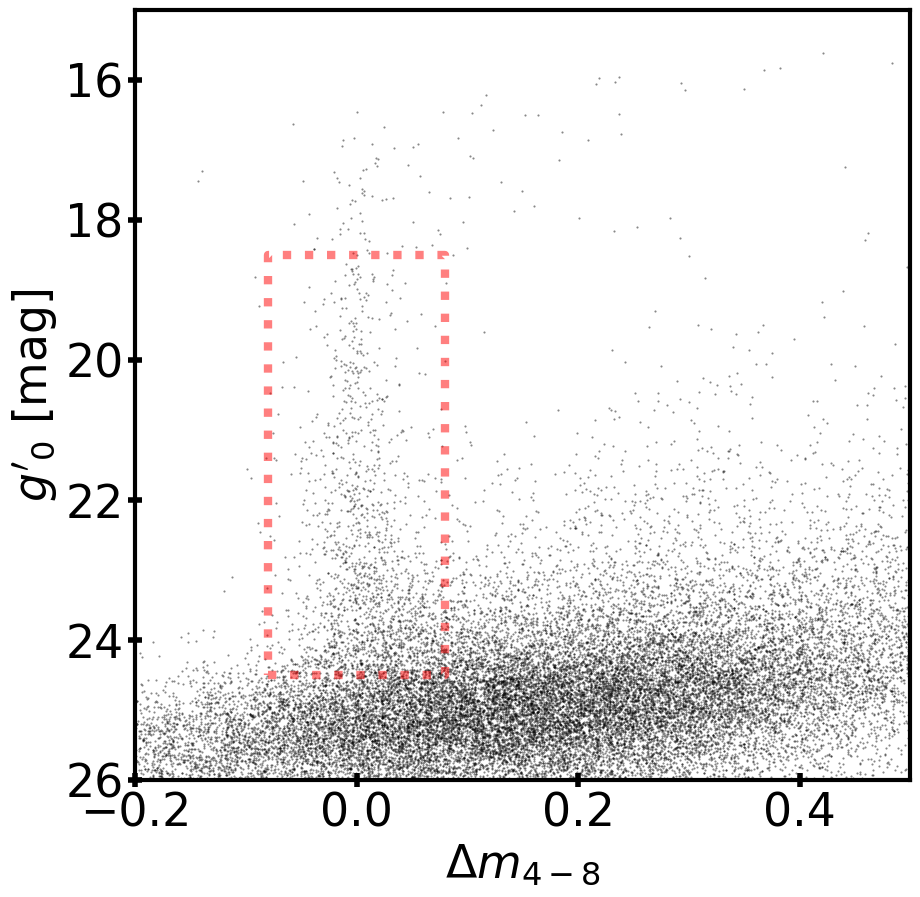}{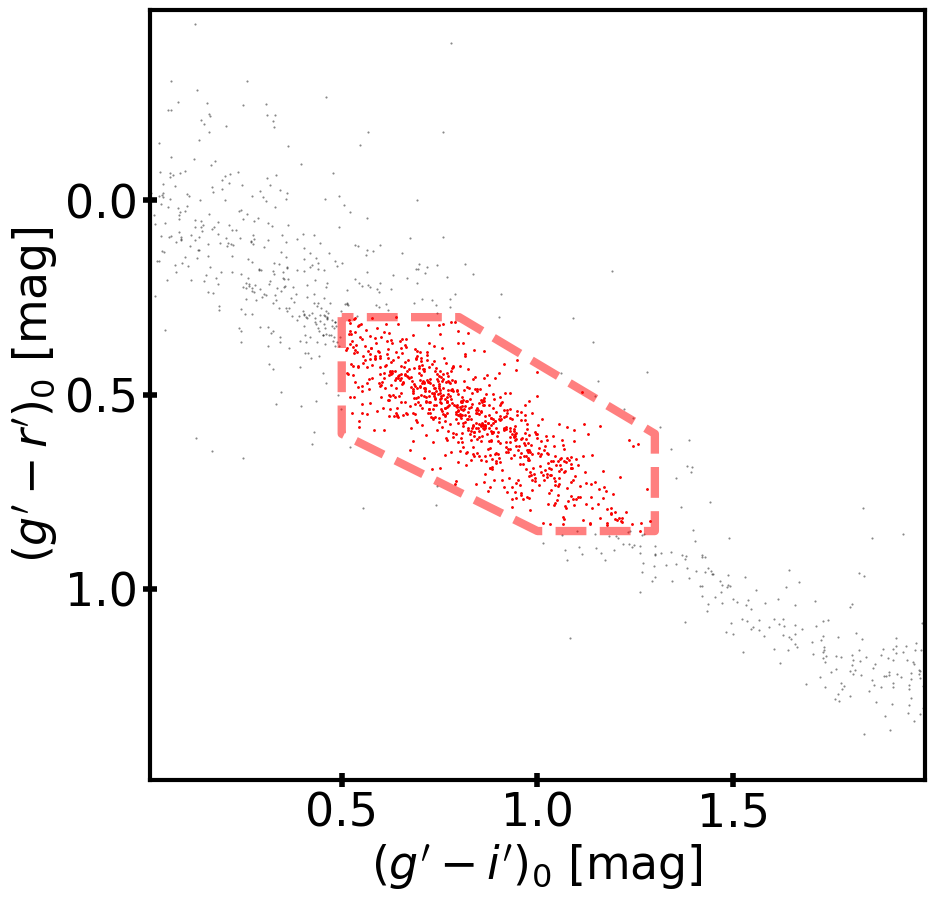}
\caption{(Left) Inverse concentration index, $\Delta m_{4-8}$, versus g-band magnitude for sources in the NGC524 region. The red dotted box shows the point-like source region used for this galaxy. (Right) ($g'-r'$)-($g'-i'$) color-color diagram of point-like sources in the NGC524 region. The red dashed polygon shows the GC selection region used in this study, with red sources showing GC candidates.
\label{gcselection2}}
\end{figure}

We selected GC candidates based on a combination of size information and colors. Since most GCs at the distances of our target galaxies appear as point-like, or slightly extended, sources in the images, we chose point-like sources first based on the inverse concentration index, $\Delta m'_{4-8}$. This is the difference in magnitudes between apertures of 4-pixel and 8-pixel diameters. These aperture magnitudes are corrected for missing point source fluxes, so the $\Delta m'_{4-8}$ value of point sources is defined to be zero. We measured $\Delta m'_{4-8}$ values using the g'- and i'-band images for the NGVS targets and combined them with error-weighting. As for the MATLAS targets, we calculated $\Delta m_{4-8}$ values on the best seeing filter image and the second best seeing images. These $\Delta m_{4-8}$ values are also combined with error-weighting. We used these error-weighted mean $\Delta m_{4-8}$ values for the point source selection. The $\Delta m_{4-8}$ values of point sources show scatter with a mean of zero due to photometric errors. Therefore, we chose point-like sources with a range of $\Delta m_{4-8}$ values, $-0.08 \leq \Delta m_{4-8} \leq 0.08$. We limited the selection to sources brighter than $g'=24.5$~mag to mitigate the effect of large photometric errors. GCs at smaller galactocentric distances can be partially resolved in high-quality images, so we expanded the range of IC values as $-0.08 \leq \Delta m_{4-8} \leq 0.16$ for galaxies within $20$~Mpc for the NGVS targets and for galaxies within $20$~Mpc in MATLAS having high image quality (Seeing $\leq 1.0 \arcsec$). After selecting point-like sources, we used color information to choose GCs. We chose GC candidates using polygons in the $(u^*-g')-(g'-i')$ color-color diagrams (Figure~\ref{gcselection}) when u-band data is available. Otherwise, we used $(g'-r')-(g'-i')$ color-color diagrams (Figure~ \ref{gcselection2}). All GC selection polygons are defined based on the M87 spectroscopically confirmed GCs (see \citealp{2017ApJ...835..123L}).

\subsection{ACS Virgo Cluster Survey data}

Although model-subtracted images allow us to detect additional sources in the central regions of galaxies, the GC samples will still be incomplete in the cores of bright galaxies. 
Because we have targeted galaxies in the ACSVCS, we also have HST/ACS photometric catalogues that are much more complete than is possible with ground-based imaging. 
Our analysis thus uses the GC catalog from ACSVCS \citep{2007ApJS..171..101J} which provides $g_{\rm ACS}$ and $z_{\rm ACS}$ magnitudes, and GC probability.
We transformed these HST magnitudes to CFHT $g$ and $i$ magnitudes using the following linear equations derived from matched sources in the M87 region. 
\begin{equation}
\label{gacs}
g'_{\rm NGVS} = g_{\rm ACS} -(0.060 \pm 0.005 ) - (0.057\pm0.004) \times (g_{\rm ACS} - z_{\rm ACS})
\end{equation}
\begin{equation}
\label{gzacs}
(g'_{\rm NGVS}-i'_{\rm NGVS}) =(g_{\rm ACS}-z_{\rm ACS}+(0.27\pm0.02))/(1.65\pm0.02)
\end{equation}
where $g'_{\rm NGVS}$, $i'_{\rm NGVS}$, $g_{\rm ACS}$, and $z_{\rm ACS}$ represent magnitudes of CFHT $g'$, CFHT $i'$, HST/ACS $F475W$, and HST/ACS $F850LP$, respectively.

For the ACSVCS sources, we define GCs with ${\rm p}_{GC} \geq 0.5$, and $g'_{\rm NGVS} \leq24.5$ mag, where ${\rm p}_{GC}$ is the probability that the object is a GC. We matched these GC candidates to those from the NGVS catalog, and those without a match are added to the total GC samples for further analysis.

\subsection{Completeness test}
To estimate the limit of our ground-based photometry, we performed completeness tests by injecting artificial stars into the $10\arcmin\times10\arcmin$ cutout images. Artificial stars were constructed using Point-Spread Functions (PSFs) empirically obtained for each observing field. PSFs were generated with DAOphot, and PSF stars were selected using SExtractor, with inverse concentration indices applied for point source selection. Detailed information on PSF generation for the NGVS is available on the NGVS webpage, and the process for generating PSFs for MATLAS data closely follows that of the NGVS. When adding these artificial stars, we used a power-law luminosity function with a magnitude range of $20 < g' < 25$. Our GC survey is limited to $g'_0 = 24.5$, so we set the faint magnitude limit to $g'=25$. When performing these tests, the number of added artificial stars did not exceed $10\%$ of the number of sources with the same magnitude range. We repeated this experiment over a thousand times, resulting in more than $150,000$ artificial stars. On average, we added about $200,000$ artificial stars to each target image. As expected, the completeness varied with both the magnitude of sources and the background brightness, which generally varies radially from the center of galaxies for ETGs. To account for this  variation, we divided the artificial star test into subgroups based on the radial distance from galaxy center and fitted the results with a step function as follows:
\begin{equation}
\label{comp}
f(m|m_{50},\alpha) = \frac{1}{2} \times \left( 1 - \frac{\alpha (m - m_{50})}{\sqrt{1+\alpha^2(m-m_{50})^2 }} \right)
\end{equation}
where $\alpha$ is the slope of the decreasing recovery rate; $m_{50}$ is the magnitude for the $50\%$ recovery rate; and $m$ is the magnitude of sources. By applying this method, we obtained the recovery rates that vary with source magnitudes and distances from the galaxy center.

We found that the ground-based photometry suffers from incompleteness near center of galaxies for the sources with $g'\sim24.5$~mag, whereas the ACSVCS catalog is $100\%$ complete for the sources with $g'_0\leq24.5$~mag in the entire fields.

\subsection{GC Density Profile Fitting}

We used an analytic function to investigate the spatial distributions of GCs, especially the GC number density profiles. We fitted these profiles using a two-dimensional S\'ersic function plus a constant background, given by:

\begin{equation}
\label{sersic}
\Sigma(R)=\Sigma_e  \exp \left\{ -b_n \left[ \left( \frac{R}{R_e} \right)^{1/n} -1 \right] \right\} + \Sigma_b
\end{equation}
where 
\begin{eqnarray}
R & = & \sqrt{(X^{\prime 2} + Y^{\prime 2})/(1-\epsilon^2)} \\
X^{\prime} & = & (X-X_0)\cos{\theta} + (Y-Y_0)\sin{\theta} \\
Y^{\prime} & = & (Y-Y_0)\cos{\theta} - (X-X_0)\sin{\theta}.
\end{eqnarray}

Here $\theta$ is the position angle of the GC candidate, measured from north to east, while $X_0$ and $Y_0$ are the coordinates of the center of the host galaxy. $X$ and $Y$ are the coordinates of the GC candidates, and $\epsilon$ is the ellipticity. In the S\'ersic function, $R_e$ is the effective radius, $\Sigma_e$ is the GC number density at the effective radius, $\Sigma_b$ is the background GC number density, $n$ is the S\'ersic index, and $b_n$ is a constant that depends on $n$. 
We used an approximation of $b_n$ \citep{1999A&A...352..447C}. 

We fitted this function to the data using a Markov Chain Monte Carlo (MCMC) method with the {\tt emcee} code in python \citep{2013PASP..125..306F}. We set flat prior distributions on several parameters, including $\Sigma_e > 0.001$~arcmin$^{-2}$, $0.25 < n < 8.0$, and $0.05<R_e<30$ arcmin. Due to difficulties in constraining both ellipticity and position angle in the presence of GC contamination, we imposed priors of $0 \leq \epsilon < 0.1$ and $-10^{\circ} < \theta < 10^{\circ}$ for most cases. For a handful of elongated galaxies, we also imposed priors of $0 \leq \epsilon < 0.4$ and $-180^{\circ} < \theta < 180^{\circ}$. We adopted a prior from a Gaussian function with pre-estimated values of the background mean  and background standard deviation for the background level.

To obtain the logarithmic probability, we used the following likelihood function:
\begin{equation}
\label{likehood}
\mathcal{L}(\Sigma_e,n,R_e,\Sigma_b) \propto \prod_i \ell_i (R_i | \Sigma_e,n,R_e,\Sigma_b)
\end{equation}
Here, $\ell_i (R_i | \Sigma_e ,n , R_e, \Sigma_n)$ is the probability of finding datum $i$ at radius $R_i$ given the S\'ersic parameters. 
We applied a completeness correction to each data point $R_i$ for galaxies without HST data. 
This completeness correction involved convolving the source detection probability function for CFHT/MegaCam data with the intrinsic luminosity function, which allowed the density probability function to be corrected for each data point.

To construct the probability function, we need to integrate the modified S\'ersic function.
As our modified S\'ersic function does not have a theoretical integrated function form, we used a numerical integration method to generate the probability function. The integration range is normally from the radius corresponding to the 50\% completeness limit to 30$^\prime$. If ACSVCS imaging is available, the integration starts at the galaxy center. For the largest galaxies, the outer limit was taken to be 60$^\prime$.

We also masked unfavourable areas for GC density profile fitting: i.e., those regions affected by saturated stars, nearby galaxies, or outside of the observation fields, etc.  
In the case of target galaxies with neighbors, we fit two 2D S\'ersic functions, simultaneously, to include the contribution from the the neighboring galaxy.
However, if there are more than two neighboring target galaxies, then we masked them except for the most dominant neighbor. 

\subsection{Color distribution of GCs}
 
The colors of GCs in early-type galaxies are often used as a metallicity indicator, and they typically show unimodal or bimodal distributions. We used Gaussian Mixture Modelling (GMM, \citealp{2010ApJ...718.1266M}) to test for color-bimodality among the GC systems. To create input catalogs for the GMM test, we selected GCs within a radius of $2.5 R_{e,gc}$, and used their $(g-i)$ colors. However, these GCs can be contaminated by various sources, including intra-cluster GCs, foreground stars, and background galaxies. To deal with background contamination, we followed several steps: (1) We defined a background area outside the galaxy region and calculated the areal fraction between the background field and the galaxy field (within $2.5 R_{e,gc}$); we then randomly chose background objects based on this areal fraction; (2) We subtracted GCs in the galaxy region that had the closest colors to the GC candidates in the background region; (3) Using this background-subtracted catalog, we ran the GMM code and obtained the results; (4) We repeated this process 30 times to account for errors in background subtraction. After these steps, we calculated the mean and standard deviation of D-values, representing the distances between the peaks of the fitted Gaussian functions. If the D-value was greater than 2, it indicates the data was better fit by two Gaussian functions, and we considered the distribution bimodal when the D-values were greater than 2 standard deviations. For GCs with a D-value greater than 2, we divided them into blue and red GCs based on the crossover values of the two fitted Gaussian functions.

\subsection{Total number of GCs}
The total number of GCs is a fundamental parameter in the study of GC systems. However, estimating this number requires overcoming the observational limits of magnitudes and spatial coverage. In this study, we used the 2D-S\'ersic fitting of the GC distribution to estimate the total number of GCs above the magnitude limit ($g'_0=24.5$~mag). We integrated the fitted 2D-S\'ersic function to obtain the magnitude-limited total number of GCs.

It is well known that GC luminosity functions (GCLFs) have a roughly universal Gaussian form with a peak luminosity of $M_V \sim -7.5$ (e.g., \citealp{2001segc.book..223H,2003JKAS...36..189L}). More precisely, peaks and widths of GCLFs change slightly depending on the host galaxy luminosity (e.g., \citealp{2007ApJS..171..101J,2010ApJ...717..603V}). We made a similar assumption by adopting different peak luminosities and widths of the Gaussian function based on the host galaxy luminosity, using the relation from \citet{2010ApJ...717..603V}. \citet{2010ApJ...717..603V} also provides mean and sigma values for the GCLF of many target galaxies in this study, so we used these values directly for these common targets. With these GCLFs, we could correct the observation's magnitude limit and obtain the total number of GCs. We were also able to estimate the errors of the GC total numbers based on the errors of the S\'ersic fitting, but we did not include the errors from GCLF. 
Note that the total number of GCs for ACS-VCS galaxies has been calculated based on the HST data \citep{2008ApJ...681..197P}; we will compare both numbers and discuss them in the next section. 

\section{Results and Discussions}

\subsection{GC properties}
Table \ref{tbl:GCsystems} presents the estimated properties of the GC systems in our target galaxies. The table provides information on the effective radii and total numbers of GC systems with their $1\sigma$ errors. Additionally, we include the S\'ersic indices, $n$, for all GC systems. The table also notes the bimodality of GC colors, indicating the effective radii and peak colors of the blue and red GCs for those galaxies that exhibit a bimodal GC color distribution. The table also includes the median GC colors for those galaxies exhibiting an unimodal GC color distribution. Furthermore, the GC specific frequency is also noted (see the bottom of this section).

\startlongtable
\begin{deluxetable*}{ccccccccccc}
\tablenum{2}
\tablecaption{GC properties from this study. Column 1 lists the names of the galaxies. Columns 2 and 3 present the effective radii and S\'ersic indices ($n$) of the GC systems, respectively. Column 4 shows the total number of GCs, and column 5 indicates the status of bimodality in the GC color distributions. Columns 6 to 9 detail the properties of GC systems with bimodal color distributions: effective radii of blue and red GC systems (columns 6 and 7), and peak colors of blue and red GC systems (columns 8 and 9). For unimodal GC color distributions, column 9 lists their median GC colors. Finally, column 11 shows the GC specific frequencies. \label{tbl:GCsystems}}
\tablewidth{0pt}
\tablehead{
\colhead{Name} & \colhead{$R_{e,gc}$} & \colhead{S\'ersic $n$} & \colhead{$N_{GC}$} & \colhead{Bimodality} & \colhead{$R_{e,bgc}$} & \colhead{$R_{e,rgc}$} & \colhead{$(g-i)_{0,b}$} & \colhead{$(g-i)_{0,r}$} & \colhead{$(g-i)_{0,m}$} & \colhead{$S_{N,g'}$} \\
\colhead{} & \colhead{[$\arcmin$] } & \colhead{} & \colhead{[$\#$]} & \colhead{} & \colhead{[$\arcmin$]} & \colhead{[$\arcmin$]} & \colhead{} & \colhead{} & \colhead{} & \colhead{}}
\decimalcolnumbers
\startdata
NGC0524 & $2.69^{+0.39}_{-0.24}$ & $2.58^{+0.90}_{-0.56}$ & $1459^{+277}_{-211}$ & N & $...$ & $...$ & $...$ & $...$ & $0.86$ & $6.9^{+1.3}_{-1.0}$ \\
NGC0821 & $1.17^{+0.43}_{-0.28}$ & $5.38^{+1.73}_{-1.90}$ & $764^{+315}_{-230}$ & Y & $1.51^{+0.70}_{-0.43}$ & $0.53^{+0.27}_{-0.15}$ & $0.76$ & $1.01$ & $...$ & $4.3^{+1.8}_{-1.3}$ \\
NGC0936 & $2.22^{+1.24}_{-0.68}$ & $6.71^{+0.93}_{-1.40}$ & $1265^{+452}_{-413}$ & Y & $4.47^{+2.80}_{-1.55}$ & $1.12^{+0.52}_{-0.43}$ & $0.74$ & $1.00$ & $...$ & $4.3^{+1.6}_{-1.4}$ \\
NGC1023 & $1.72^{+0.60}_{-0.43}$ & $4.12^{+2.34}_{-2.28}$ & $304^{+104}_{-93}$ & Y & $2.05^{+1.23}_{-0.75}$ & $1.55^{+0.32}_{-0.44}$ & $0.67$ & $0.92$ & $...$ & $2.5^{+0.9}_{-0.8}$ \\
NGC2592 & $0.73^{+0.31}_{-0.20}$ & $4.90^{+2.07}_{-1.96}$ & $269^{+72}_{-67}$ & N & $...$ & $...$ & $...$ & $...$ & $0.81$ & $6.3^{+1.7}_{-1.6}$ \\
NGC2685 & $1.65^{+0.78}_{-0.75}$ & $2.85^{+3.57}_{-2.12}$ & $50^{+28}_{-19}$ & Y & $1.35^{+0.83}_{-0.69}$ & $0.61^{+0.87}_{-0.33}$ & $0.80$ & $1.10$ & $...$ & $1.1^{+0.6}_{-0.4}$ \\
NGC2768 & $1.45^{+0.46}_{-0.37}$ & $5.58^{+1.59}_{-1.86}$ & $1012^{+307}_{-225}$ & Y & $1.78^{+0.54}_{-0.37}$ & $0.46^{+0.24}_{-0.13}$ & $0.73$ & $0.98$ & $...$ & $4.4^{+1.3}_{-1.0}$ \\
NGC2778 & $1.54^{+0.88}_{-0.69}$ & $5.20^{+2.04}_{-2.29}$ & $94^{+50}_{-36}$ & Y & $2.92^{+1.84}_{-1.34}$ & $0.26^{+0.23}_{-0.13}$ & $0.77$ & $1.14$ & $...$ & $3.0^{+1.6}_{-1.1}$ \\
NGC2950 & $0.67^{+0.34}_{-0.20}$ & $3.21^{+3.11}_{-1.83}$ & $83^{+31}_{-26}$ & Y & $0.86^{+0.59}_{-0.31}$ & $0.58^{+0.36}_{-0.24}$ & $0.66$ & $0.91$ & $...$ & $1.4^{+0.5}_{-0.4}$ \\
NGC3098 & $1.51^{+1.11}_{-0.58}$ & $4.72^{+2.22}_{-2.45}$ & $123^{+60}_{-39}$ & N & $...$ & $...$ & $...$ & $...$ & $0.79$ & $2.5^{+1.2}_{-0.8}$ \\
NGC3245 & $0.85^{+0.37}_{-0.24}$ & $5.10^{+1.77}_{-1.94}$ & $325^{+88}_{-82}$ & Y & $1.23^{+0.40}_{-0.32}$ & $0.36^{+0.20}_{-0.11}$ & $0.74$ & $1.04$ & $...$ & $3.0^{+0.8}_{-0.8}$ \\
NGC3379 & $1.97^{+0.06}_{-0.07}$ & $7.97^{+0.02}_{-0.05}$ & $352^{+22}_{-26}$ & Y & $2.02^{+0.06}_{-0.03}$ & $0.84^{+0.06}_{-0.06}$ & $0.77$ & $1.03$ & $...$ & $3.3^{+0.2}_{-0.2}$ \\
NGC3384 & $0.93^{+0.10}_{-0.10}$ & $0.52^{+0.17}_{-0.16}$ & $50^{+10}_{-9}$ & Y & $1.09^{+0.03}_{-0.06}$ & $1.08^{+0.05}_{-0.05}$ & $0.77$ & $0.97$ & $...$ & $0.6^{+0.1}_{-0.1}$ \\
NGC3457 & $0.34^{+0.13}_{-0.07}$ & $1.56^{+3.45}_{-1.01}$ & $39^{+18}_{-13}$ & N & $...$ & $...$ & $...$ & $...$ & $0.85$ & $1.5^{+0.7}_{-0.5}$ \\
NGC3489 & $1.07^{+0.68}_{-0.37}$ & $4.84^{+2.23}_{-2.46}$ & $114^{+35}_{-36}$ & Y & $0.83^{+0.51}_{-0.29}$ & $0.86^{+1.10}_{-0.49}$ & $0.73$ & $1.19$ & $...$ & $1.6^{+0.5}_{-0.5}$ \\
NGC3599 & $0.83^{+0.44}_{-0.31}$ & $3.69^{+2.83}_{-2.15}$ & $97^{+35}_{-27}$ & N & $...$ & $...$ & $...$ & $...$ & $0.80$ & $2.4^{+0.9}_{-0.7}$ \\
NGC3607 & $2.03^{+0.09}_{-0.07}$ & $7.97^{+0.03}_{-0.07}$ & $865^{+77}_{-61}$ & Y & $1.70^{+0.06}_{-0.04}$ & $1.70^{+0.08}_{-0.07}$ & $0.81$ & $1.05$ & $...$ & $2.9^{+0.3}_{-0.2}$ \\
NGC3608 & $2.01^{+0.13}_{-0.11}$ & $1.56^{+0.30}_{-0.21}$ & $376^{+60}_{-37}$ & Y & $1.14^{+0.15}_{-0.12}$ & $0.96^{+0.06}_{-0.08}$ & $0.76$ & $1.03$ & $...$ & $3.8^{+0.6}_{-0.4}$ \\
NGC3630 & $1.82^{+2.09}_{-0.85}$ & $6.83^{+0.83}_{-1.93}$ & $305^{+227}_{-118}$ & N & $...$ & $...$ & $...$ & $...$ & $0.73$ & $4.6^{+3.4}_{-1.8}$ \\
NGC3945 & $1.75^{+1.06}_{-0.52}$ & $2.67^{+3.02}_{-1.98}$ & $119^{+66}_{-47}$ & Y & $2.07^{+1.07}_{-0.74}$ & $0.95^{+0.63}_{-0.40}$ & $0.70$ & $1.07$ & $...$ & $0.6^{+0.3}_{-0.2}$ \\
IC3032 & $0.45^{+0.27}_{-0.19}$ & $2.06^{+2.36}_{-1.24}$ & $9^{+6}_{-4}$ & N & $...$ & $...$ & $...$ & $...$ & $0.86$ & $3.8^{+2.6}_{-1.7}$ \\
IC3065 & $0.70^{+0.50}_{-0.32}$ & $2.81^{+2.18}_{-1.28}$ & $45^{+17}_{-14}$ & N & $...$ & $...$ & $...$ & $...$ & $0.76$ & $6.4^{+2.4}_{-2.0}$ \\
VCC200 & $0.28^{+0.08}_{-0.07}$ & $0.80^{+1.28}_{-1.28}$ & $24^{+10}_{-8}$ & N & $...$ & $...$ & $...$ & $...$ & $0.68$ & $5.1^{+2.1}_{-1.6}$ \\
IC3101 & $0.29^{+0.07}_{-0.05}$ & $3.13^{+0.91}_{-1.33}$ & $31^{+14}_{-10}$ & N & $...$ & $...$ & $...$ & $...$ & $0.71$ & $13.4^{+6.1}_{-4.3}$ \\
NGC4262 & $1.69^{+0.65}_{-0.49}$ & $2.58^{+0.99}_{-0.87}$ & $154^{+46}_{-38}$ & Y & $2.76^{+1.40}_{-0.93}$ & $0.58^{+0.30}_{-0.19}$ & $0.70$ & $1.07$ & $...$ & $3.9^{+1.2}_{-1.0}$ \\
NGC4267 & $0.97^{+0.05}_{-0.05}$ & $0.46^{+0.12}_{-0.08}$ & $229^{+42}_{-38}$ & Y & $0.97^{+0.21}_{-0.14}$ & $0.97^{+0.06}_{-0.06}$ & $0.75$ & $0.95$ & $...$ & $3.1^{+0.6}_{-0.5}$ \\
NGC4278 & $2.53^{+0.11}_{-0.08}$ & $3.12^{+0.16}_{-0.09}$ & $1188^{+98}_{-73}$ & Y & $1.22^{+0.03}_{-0.08}$ & $1.15^{+0.06}_{-0.07}$ & $0.79$ & $1.02$ & $...$ & $10.3^{+0.8}_{-0.6}$ \\
NGC4283 & $1.00^{+0.12}_{-0.10}$ & $7.99^{+0.01}_{-0.02}$ & $192^{+54}_{-38}$ & Y & $3.01^{+0.06}_{-0.04}$ & $0.21^{+0.03}_{-0.01}$ & $0.71$ & $1.02$ & $...$ & $10.0^{+2.8}_{-2.0}$ \\
UGC7436 & $0.42^{+0.23}_{-0.16}$ & $5.84^{+1.50}_{-1.97}$ & $38^{+15}_{-12}$ & N & $...$ & $...$ & $...$ & $...$ & $0.72$ & $6.1^{+2.3}_{-1.8}$ \\
VCC571 & $0.12^{+0.12}_{-0.05}$ & $5.16^{+1.99}_{-2.70}$ & $19^{+9}_{-7}$ & N & $...$ & $...$ & $...$ & $...$ & $0.75$ & $3.0^{+1.4}_{-1.2}$ \\
NGC4318 & $0.47^{+0.35}_{-0.19}$ & $6.68^{+0.97}_{-2.21}$ & $43^{+28}_{-19}$ & Y & $0.22^{+0.32}_{-0.10}$ & $0.46^{+0.40}_{-0.22}$ & $0.66$ & $0.89$ & $...$ & $2.6^{+1.7}_{-1.1}$ \\
NGC4339 & $1.34^{+0.17}_{-0.13}$ & $1.05^{+0.25}_{-0.25}$ & $223^{+30}_{-26}$ & N & $...$ & $...$ & $...$ & $...$ & $0.83$ & $4.7^{+0.6}_{-0.6}$ \\
NGC4340 & $0.76^{+0.08}_{-0.10}$ & $0.86^{+0.12}_{-0.16}$ & $76^{+19}_{-23}$ & Y & $0.83^{+0.10}_{-0.10}$ & $0.77^{+0.08}_{-0.07}$ & $0.67$ & $0.94$ & $...$ & $1.0^{+0.2}_{-0.3}$ \\
NGC4342 & $1.76^{+2.84}_{-0.10}$ & $0.73^{+4.72}_{-0.12}$ & $527^{+54}_{-107}$ & Y & $1.11^{+0.09}_{-0.06}$ & $1.07^{+0.05}_{-0.07}$ & $0.69$ & $0.82$ & $...$ & $20.6^{+2.1}_{-4.2}$ \\
NGC4350 & $2.31^{+0.16}_{-0.22}$ & $2.99^{+0.12}_{-0.13}$ & $459^{+73}_{-80}$ & Y & $2.41^{+0.16}_{-0.24}$ & $1.36^{+0.93}_{-0.53}$ & $0.73$ & $0.98$ & $...$ & $6.1^{+1.0}_{-1.1}$ \\
NGC4352 & $0.59^{+0.05}_{-0.04}$ & $0.86^{+0.30}_{-0.20}$ & $158^{+41}_{-39}$ & N & $...$ & $...$ & $...$ & $...$ & $0.74$ & $6.8^{+1.7}_{-1.7}$ \\
NGC4365 & $3.73^{+0.13}_{-0.11}$ & $2.04^{+0.17}_{-0.11}$ & $3887^{+243}_{-238}$ & Y & $4.95^{+0.31}_{-0.29}$ & $2.92^{+0.16}_{-0.14}$ & $0.72$ & $0.95$ & $...$ & $6.1^{+0.4}_{-0.4}$ \\
NGC4371 & $1.03^{+0.27}_{-0.18}$ & $2.58^{+0.92}_{-0.63}$ & $277^{+81}_{-59}$ & Y & $1.42^{+0.95}_{-0.47}$ & $0.98^{+0.28}_{-0.15}$ & $0.74$ & $1.01$ & $...$ & $2.8^{+0.8}_{-0.6}$ \\
NGC4374 & $6.18^{+0.74}_{-0.60}$ & $2.71^{+0.31}_{-0.24}$ & $3080^{+388}_{-319}$ & Y & $7.93^{+0.82}_{-0.72}$ & $3.17^{+0.36}_{-0.26}$ & $0.73$ & $0.99$ & $...$ & $4.7^{+0.6}_{-0.5}$ \\
NGC4377 & $0.42^{+0.08}_{-0.05}$ & $0.87^{+0.50}_{-0.40}$ & $83^{+24}_{-19}$ & Y & $0.58^{+0.20}_{-0.12}$ & $0.34^{+0.07}_{-0.05}$ & $0.73$ & $0.97$ & $...$ & $2.0^{+0.6}_{-0.5}$ \\
NGC4379 & $0.52^{+0.16}_{-0.09}$ & $1.78^{+1.33}_{-0.60}$ & $92^{+27}_{-18}$ & N & $...$ & $...$ & $...$ & $...$ & $0.86$ & $2.5^{+0.7}_{-0.5}$ \\
NGC4387 & $0.68^{+0.31}_{-0.19}$ & $1.66^{+1.00}_{-0.72}$ & $62^{+39}_{-24}$ & Y & $0.98^{+0.35}_{-0.36}$ & $0.35^{+0.20}_{-0.11}$ & $0.69$ & $0.91$ & $...$ & $2.0^{+1.3}_{-0.8}$ \\
IC3328 & $0.50^{+0.13}_{-0.10}$ & $2.84^{+1.35}_{-0.93}$ & $68^{+21}_{-15}$ & N & $...$ & $...$ & $...$ & $...$ & $0.77$ & $8.9^{+2.7}_{-2.0}$ \\
NGC4406 & $7.16^{+0.69}_{-0.52}$ & $1.95^{+0.19}_{-0.14}$ & $3261^{+394}_{-315}$ & Y & $8.35^{+0.96}_{-0.82}$ & $4.58^{+0.47}_{-0.36}$ & $0.71$ & $0.97$ & $...$ & $4.2^{+0.5}_{-0.4}$ \\
NGC4417 & $0.93^{+0.08}_{-0.10}$ & $1.98^{+0.18}_{-0.18}$ & $100^{+17}_{-23}$ & Y & $1.04^{+0.18}_{-0.16}$ & $0.48^{+0.06}_{-0.06}$ & $0.72$ & $0.99$ & $...$ & $1.6^{+0.3}_{-0.4}$ \\
NGC4425 & $1.71^{+0.18}_{-0.16}$ & $7.85^{+0.09}_{-0.10}$ & $213^{+58}_{-42}$ & Y & $0.60^{+0.04}_{-0.04}$ & $0.37^{+0.05}_{-0.06}$ & $0.69$ & $0.87$ & $...$ & $6.6^{+1.8}_{-1.3}$ \\
NGC4429 & $1.01^{+0.28}_{-0.16}$ & $1.94^{+2.72}_{-1.07}$ & $269^{+179}_{-92}$ & Y & $0.70^{+0.36}_{-0.26}$ & $1.20^{+0.46}_{-0.23}$ & $0.70$ & $0.97$ & $...$ & $1.4^{+0.9}_{-0.5}$ \\
NGC4434 & $0.95^{+0.21}_{-0.14}$ & $4.35^{+1.46}_{-1.46}$ & $159^{+27}_{-23}$ & N & $...$ & $...$ & $...$ & $...$ & $0.74$ & $3.0^{+0.5}_{-0.4}$ \\
NGC4435 & $0.90^{+0.09}_{-0.09}$ & $1.74^{+0.36}_{-0.32}$ & $224^{+34}_{-28}$ & Y & $1.17^{+0.20}_{-0.22}$ & $0.77^{+0.09}_{-0.07}$ & $0.73$ & $0.99$ & $...$ & $2.0^{+0.3}_{-0.2}$ \\
NGC4442 & $1.09^{+0.11}_{-0.09}$ & $1.24^{+0.30}_{-0.21}$ & $308^{+38}_{-34}$ & Y & $1.35^{+0.22}_{-0.16}$ & $0.85^{+0.10}_{-0.09}$ & $0.74$ & $1.00$ & $...$ & $3.0^{+0.4}_{-0.3}$ \\
IC3383 & $0.54^{+0.19}_{-0.11}$ & $7.92^{+0.07}_{-0.19}$ & $29^{+14}_{-11}$ & N & $...$ & $...$ & $...$ & $...$ & $0.72$ & $7.8^{+3.7}_{-2.9}$ \\
IC3381 & $1.04^{+0.37}_{-0.23}$ & $6.28^{+1.20}_{-1.63}$ & $125^{+44}_{-35}$ & N & $...$ & $...$ & $...$ & $...$ & $0.72$ & $7.5^{+2.6}_{-2.1}$ \\
NGC4452 & $2.09^{+1.13}_{-0.85}$ & $5.08^{+1.62}_{-1.58}$ & $155^{+71}_{-53}$ & Y & $1.51^{+0.83}_{-0.52}$ & $0.23^{+0.33}_{-0.11}$ & $0.70$ & $1.06$ & $...$ & $6.6^{+3.0}_{-2.2}$ \\
NGC4458 & $0.61^{+0.15}_{-0.10}$ & $2.14^{+1.16}_{-0.68}$ & $124^{+34}_{-30}$ & N & $...$ & $...$ & $...$ & $...$ & $0.93$ & $3.9^{+1.1}_{-0.9}$ \\
NGC4459 & $1.10^{+0.14}_{-0.10}$ & $1.35^{+0.32}_{-0.27}$ & $278^{+38}_{-31}$ & Y & $1.45^{+0.47}_{-0.28}$ & $0.88^{+0.11}_{-0.09}$ & $0.78$ & $1.00$ & $...$ & $1.8^{+0.2}_{-0.2}$ \\
NGC4461 & $2.12^{+0.23}_{-0.19}$ & $1.92^{+0.19}_{-0.23}$ & $288^{+62}_{-48}$ & Y & $0.67^{+0.04}_{-0.04}$ & $0.67^{+0.06}_{-0.07}$ & $0.68$ & $1.02$ & $...$ & $4.2^{+0.9}_{-0.7}$ \\
VCC1185 & $0.59^{+0.12}_{-0.17}$ & $1.38^{+0.48}_{-0.32}$ & $30^{+14}_{-12}$ & N & $...$ & $...$ & $...$ & $...$ & $0.71$ & $11.6^{+5.4}_{-4.6}$ \\
NGC4472 & $10.66^{+0.54}_{-0.43}$ & $2.45^{+0.12}_{-0.13}$ & $9826^{+839}_{-758}$ & Y & $13.95^{+0.45}_{-0.45}$ & $5.34^{+0.25}_{-0.23}$ & $0.72$ & $1.01$ & $...$ & $8.6^{+0.7}_{-0.7}$ \\
NGC4473 & $2.48^{+0.88}_{-0.45}$ & $3.14^{+0.81}_{-0.56}$ & $674^{+161}_{-132}$ & Y & $4.11^{+1.82}_{-1.11}$ & $1.40^{+0.27}_{-0.18}$ & $0.73$ & $0.98$ & $...$ & $4.6^{+1.1}_{-0.9}$ \\
NGC4474 & $0.93^{+0.24}_{-0.15}$ & $2.69^{+1.01}_{-0.67}$ & $183^{+39}_{-28}$ & N & $...$ & $...$ & $...$ & $...$ & $0.84$ & $4.5^{+1.0}_{-0.7}$ \\
NGC4476 & $0.49^{+0.14}_{-0.13}$ & $2.98^{+0.83}_{-0.83}$ & $30^{+14}_{-10}$ & N & $...$ & $...$ & $...$ & $...$ & $0.77$ & $0.9^{+0.4}_{-0.3}$ \\
NGC4477 & $1.98^{+0.10}_{-0.11}$ & $1.50^{+0.15}_{-0.16}$ & $264^{+22}_{-29}$ & Y & $2.01^{+0.10}_{-0.05}$ & $1.97^{+0.05}_{-0.05}$ & $0.74$ & $1.07$ & $...$ & $2.1^{+0.2}_{-0.2}$ \\
NGC4482 & $0.98^{+0.60}_{-0.40}$ & $3.95^{+2.20}_{-1.51}$ & $79^{+39}_{-30}$ & N & $...$ & $...$ & $...$ & $...$ & $0.76$ & $4.5^{+2.2}_{-1.8}$ \\
NGC4478 & $0.59^{+0.10}_{-0.07}$ & $2.63^{+0.89}_{-0.89}$ & $145^{+58}_{-30}$ & Y & $0.71^{+0.13}_{-0.09}$ & $0.60^{+0.16}_{-0.13}$ & $0.70$ & $0.94$ & $...$ & $2.5^{+1.0}_{-0.5}$ \\
NGC4479 & $0.77^{+0.13}_{-0.15}$ & $2.99^{+0.80}_{-0.80}$ & $49^{+18}_{-13}$ & Y & $0.47^{+0.06}_{-0.06}$ & $0.29^{+0.07}_{-0.05}$ & $0.72$ & $1.04$ & $...$ & $2.5^{+0.9}_{-0.7}$ \\
NGC4483 & $1.19^{+0.62}_{-0.42}$ & $4.64^{+1.87}_{-1.47}$ & $103^{+27}_{-24}$ & Y & $1.06^{+0.49}_{-0.33}$ & $0.08^{+0.07}_{-0.02}$ & $0.67$ & $0.90$ & $...$ & $4.2^{+1.1}_{-1.0}$ \\
NGC4486 & $13.67^{+0.88}_{-0.77}$ & $3.76^{+0.14}_{-0.17}$ & $17730^{+1030}_{-942}$ & Y & $15.22^{+1.22}_{-1.30}$ & $4.00^{+0.23}_{-0.20}$ & $0.70$ & $0.99$ & $...$ & $22.6^{+1.3}_{-1.2}$ \\
NGC4489 & $0.53^{+0.31}_{-0.11}$ & $2.04^{+2.32}_{-0.90}$ & $72^{+25}_{-19}$ & N & $...$ & $...$ & $...$ & $...$ & $0.75$ & $3.2^{+1.1}_{-0.8}$ \\
IC3461 & $0.16^{+0.02}_{-0.02}$ & $1.40^{+0.24}_{-0.20}$ & $35^{+12}_{-9}$ & Y & $0.38^{+0.08}_{-0.07}$ & $0.23^{+0.05}_{-0.05}$ & $0.71$ & $0.90$ & $...$ & $10.1^{+3.4}_{-2.7}$ \\
NGC4503 & $2.01^{+0.09}_{-0.12}$ & $3.04^{+0.23}_{-0.33}$ & $376^{+44}_{-54}$ & Y & $1.97^{+0.44}_{-0.33}$ & $0.58^{+0.05}_{-0.05}$ & $0.72$ & $1.03$ & $...$ & $5.5^{+0.7}_{-0.8}$ \\
IC3468 & $0.27^{+0.15}_{-0.08}$ & $4.44^{+2.29}_{-2.01}$ & $39^{+12}_{-10}$ & Y & $0.36^{+0.35}_{-0.15}$ & $0.25^{+0.20}_{-0.08}$ & $0.74$ & $0.89$ & $...$ & $3.0^{+0.9}_{-0.8}$ \\
IC3470 & $0.37^{+0.07}_{-0.06}$ & $1.55^{+0.82}_{-0.45}$ & $87^{+18}_{-16}$ & Y & $0.36^{+0.04}_{-0.03}$ & $0.42^{+0.28}_{-0.18}$ & $0.76$ & $0.99$ & $...$ & $15.1^{+3.1}_{-2.7}$ \\
IC798 & $0.45^{+0.11}_{-0.09}$ & $7.89^{+0.10}_{-0.34}$ & $61^{+22}_{-22}$ & N & $...$ & $...$ & $...$ & $...$ & $0.69$ & $14.6^{+5.3}_{-5.4}$ \\
NGC4515 & $0.56^{+0.27}_{-0.11}$ & $2.23^{+1.67}_{-0.82}$ & $115^{+38}_{-23}$ & N & $...$ & $...$ & $...$ & $...$ & $0.73$ & $6.2^{+2.1}_{-1.2}$ \\
VCC1512 & $0.31^{+0.40}_{-0.14}$ & $3.99^{+2.75}_{-2.75}$ & $14^{+10}_{-6}$ & N & $...$ & $...$ & $...$ & $...$ & $0.64$ & $5.7^{+4.4}_{-2.6}$ \\
IC3501 & $0.56^{+0.20}_{-0.12}$ & $2.40^{+1.30}_{-0.81}$ & $63^{+17}_{-14}$ & N & $...$ & $...$ & $...$ & $...$ & $0.77$ & $11.8^{+3.3}_{-2.7}$ \\
NGC4528 & $1.55^{+1.40}_{-0.60}$ & $2.44^{+1.43}_{-1.01}$ & $83^{+41}_{-27}$ & N & $...$ & $...$ & $...$ & $...$ & $2.18$ & $3.1^{+1.5}_{-1.0}$ \\
VCC1539 & $0.15^{+0.03}_{-0.03}$ & $1.24^{+2.19}_{-0.63}$ & $45^{+13}_{-11}$ & N & $...$ & $...$ & $...$ & $...$ & $0.76$ & $29.2^{+8.4}_{-6.9}$ \\
IC3509 & $0.47^{+0.07}_{-0.06}$ & $2.46^{+0.45}_{-0.57}$ & $71^{+20}_{-14}$ & Y & $0.44^{+0.12}_{-0.09}$ & $0.42^{+0.17}_{-0.13}$ & $0.61$ & $0.81$ & $...$ & $20.9^{+5.8}_{-4.0}$ \\
NGC4550 & $0.77^{+0.05}_{-0.05}$ & $2.32^{+0.07}_{-0.06}$ & $97^{+14}_{-10}$ & Y & $0.62^{+0.03}_{-0.02}$ & $0.49^{+0.03}_{-0.04}$ & $0.67$ & $0.92$ & $...$ & $2.7^{+0.4}_{-0.3}$ \\
NGC4551 & $0.62^{+0.08}_{-0.08}$ & $3.70^{+0.07}_{-0.07}$ & $68^{+19}_{-15}$ & Y & $0.54^{+0.08}_{-0.06}$ & $0.28^{+0.05}_{-0.05}$ & $0.72$ & $1.01$ & $...$ & $2.3^{+0.6}_{-0.5}$ \\
NGC4552 & $4.26^{+0.57}_{-0.46}$ & $3.36^{+0.38}_{-0.33}$ & $1822^{+174}_{-143}$ & Y & $6.35^{+1.52}_{-1.16}$ & $2.39^{+0.38}_{-0.26}$ & $0.72$ & $0.99$ & $...$ & $6.8^{+0.7}_{-0.5}$ \\
VCC1661 & $0.27^{+0.06}_{-0.05}$ & $7.87^{+0.10}_{-0.33}$ & $26^{+12}_{-9}$ & N & $...$ & $...$ & $...$ & $...$ & $0.72$ & $18.2^{+8.1}_{-6.5}$ \\
NGC4564 & $0.76^{+0.14}_{-0.08}$ & $2.10^{+0.61}_{-0.45}$ & $218^{+69}_{-63}$ & Y & $1.11^{+0.65}_{-0.38}$ & $0.68^{+0.09}_{-0.07}$ & $0.69$ & $0.94$ & $...$ & $3.3^{+1.0}_{-1.0}$ \\
NGC4570 & $1.33^{+0.24}_{-0.19}$ & $2.37^{+0.69}_{-0.46}$ & $261^{+70}_{-73}$ & Y & $1.76^{+0.61}_{-0.35}$ & $0.57^{+0.10}_{-0.08}$ & $0.70$ & $1.01$ & $...$ & $2.7^{+0.7}_{-0.8}$ \\
NGC4578 & $1.15^{+1.09}_{-0.36}$ & $2.53^{+1.67}_{-1.38}$ & $90^{+49}_{-29}$ & Y & $1.49^{+0.59}_{-0.61}$ & $0.66^{+0.19}_{-0.12}$ & $0.70$ & $1.06$ & $...$ & $1.8^{+1.0}_{-0.6}$ \\
NGC4596 & $2.14^{+0.98}_{-0.62}$ & $6.45^{+1.10}_{-1.70}$ & $1010^{+212}_{-210}$ & Y & $2.83^{+1.30}_{-0.85}$ & $1.10^{+0.63}_{-0.38}$ & $0.73$ & $0.99$ & $...$ & $6.6^{+1.4}_{-1.4}$ \\
VCC1826 & $0.79^{+0.62}_{-0.36}$ & $2.06^{+2.06}_{-1.23}$ & $14^{+11}_{-8}$ & N & $...$ & $...$ & $...$ & $...$ & $0.72$ & $8.1^{+6.3}_{-4.4}$ \\
VCC1833 & $0.34^{+0.19}_{-0.12}$ & $3.52^{+2.86}_{-1.87}$ & $28^{+12}_{-9}$ & N & $...$ & $...$ & $...$ & $...$ & $0.74$ & $5.6^{+2.4}_{-1.8}$ \\
IC3647 & $0.34^{+0.05}_{-0.05}$ & $0.63^{+0.85}_{-0.30}$ & $18^{+7}_{-4}$ & Y & $0.33^{+0.05}_{-0.04}$ & $0.42^{+0.35}_{-0.19}$ & $0.70$ & $0.94$ & $...$ & $3.1^{+1.1}_{-0.8}$ \\
IC3652 & $0.77^{+0.13}_{-0.12}$ & $1.89^{+0.63}_{-0.63}$ & $60^{+20}_{-15}$ & Y & $0.92^{+0.14}_{-0.14}$ & $0.43^{+0.11}_{-0.10}$ & $0.69$ & $0.99$ & $...$ & $8.0^{+2.6}_{-2.1}$ \\
NGC4608 & $3.45^{+0.93}_{-0.89}$ & $3.62^{+2.84}_{-1.97}$ & $287^{+130}_{-94}$ & Y & $3.40^{+1.01}_{-1.09}$ & $0.38^{+0.48}_{-0.24}$ & $0.71$ & $1.08$ & $...$ & $3.8^{+1.7}_{-1.2}$ \\
IC3653 & $0.41^{+0.11}_{-0.10}$ & $2.12^{+0.31}_{-0.31}$ & $14^{+9}_{-6}$ & N & $...$ & $...$ & $...$ & $...$ & $0.71$ & $2.3^{+1.4}_{-1.0}$ \\
NGC4612 & $0.88^{+0.23}_{-0.17}$ & $2.34^{+0.85}_{-0.60}$ & $159^{+50}_{-41}$ & N & $...$ & $...$ & $...$ & $...$ & $0.81$ & $2.5^{+0.8}_{-0.7}$ \\
VCC1886 & $0.25^{+0.10}_{-0.06}$ & $1.17^{+0.49}_{-0.49}$ & $7^{+6}_{-3}$ & N & $...$ & $...$ & $...$ & $...$ & $0.77$ & $3.0^{+2.7}_{-1.3}$ \\
UGC7854 & $0.97^{+0.22}_{-0.17}$ & $0.50^{+0.42}_{-0.20}$ & $18^{+10}_{-6}$ & N & $...$ & $...$ & $...$ & $...$ & $0.75$ & $5.7^{+3.0}_{-1.8}$ \\
NGC4621 & $4.48^{+0.72}_{-0.56}$ & $3.26^{+0.33}_{-0.31}$ & $1318^{+132}_{-110}$ & Y & $5.90^{+0.88}_{-0.83}$ & $3.29^{+0.41}_{-0.39}$ & $0.73$ & $0.98$ & $...$ & $5.2^{+0.5}_{-0.4}$ \\
NGC4638 & $1.62^{+0.35}_{-0.22}$ & $1.31^{+0.52}_{-0.52}$ & $234^{+62}_{-51}$ & N & $...$ & $...$ & $...$ & $...$ & $0.74$ & $3.3^{+0.9}_{-0.7}$ \\
NGC4649 & $15.71^{+0.41}_{-0.38}$ & $4.57^{+0.16}_{-0.16}$ & $8875^{+508}_{-419}$ & Y & $13.88^{+1.09}_{-1.05}$ & $3.47^{+0.18}_{-0.17}$ & $0.70$ & $0.99$ & $...$ & $14.2^{+0.8}_{-0.7}$ \\
VCC1993 & $0.65^{+0.47}_{-0.24}$ & $1.10^{+1.02}_{-0.63}$ & $6^{+7}_{-3}$ & N & $...$ & $...$ & $...$ & $...$ & $0.67$ & $2.6^{+2.9}_{-1.4}$ \\
NGC4660 & $1.06^{+0.24}_{-0.14}$ & $2.54^{+0.66}_{-0.48}$ & $291^{+50}_{-36}$ & N & $...$ & $...$ & $...$ & $...$ & $0.76$ & $5.5^{+0.9}_{-0.7}$ \\
IC3735 & $0.45^{+0.34}_{-0.17}$ & $4.18^{+2.45}_{-1.92}$ & $28^{+14}_{-10}$ & N & $...$ & $...$ & $...$ & $...$ & $0.76$ & $4.5^{+2.3}_{-1.6}$ \\
IC3773 & $0.32^{+0.20}_{-0.11}$ & $3.73^{+2.69}_{-2.22}$ & $19^{+9}_{-6}$ & N & $...$ & $...$ & $...$ & $...$ & $0.72$ & $2.9^{+1.3}_{-1.0}$ \\
IC3779 & $0.27^{+0.10}_{-0.08}$ & $3.49^{+3.03}_{-2.06}$ & $15^{+6}_{-5}$ & N & $...$ & $...$ & $...$ & $...$ & $0.66$ & $5.7^{+2.5}_{-1.8}$ \\
NGC4694 & $1.94^{+0.40}_{-0.30}$ & $3.02^{+1.82}_{-0.98}$ & $687^{+158}_{-115}$ & Y & $1.93^{+0.65}_{-0.48}$ & $1.79^{+0.50}_{-0.38}$ & $0.65$ & $0.83$ & $...$ & $12.0^{+2.8}_{-2.0}$ \\
NGC4710 & $6.94^{+2.03}_{-1.74}$ & $2.09^{+0.69}_{-0.58}$ & $1004^{+381}_{-307}$ & Y & $7.38^{+1.75}_{-1.99}$ & $5.70^{+3.35}_{-1.52}$ & $0.69$ & $1.12$ & $...$ & $10.6^{+4.0}_{-3.2}$ \\
NGC4733 & $2.52^{+1.25}_{-0.97}$ & $2.81^{+2.29}_{-1.42}$ & $90^{+61}_{-38}$ & Y & $1.60^{+1.32}_{-0.79}$ & $2.41^{+1.01}_{-0.91}$ & $0.73$ & $1.12$ & $...$ & $3.2^{+2.2}_{-1.3}$ \\
NGC4754 & $1.07^{+0.20}_{-0.15}$ & $1.24^{+0.55}_{-0.41}$ & $130^{+29}_{-23}$ & Y & $1.64^{+0.93}_{-0.41}$ & $1.11^{+0.29}_{-0.19}$ & $0.75$ & $1.03$ & $...$ & $1.1^{+0.3}_{-0.2}$ \\
NGC4762 & $1.52^{+0.57}_{-0.29}$ & $2.35^{+0.89}_{-0.59}$ & $321^{+102}_{-75}$ & Y & $2.88^{+1.90}_{-1.06}$ & $0.99^{+0.19}_{-0.14}$ & $0.67$ & $0.90$ & $...$ & $1.4^{+0.4}_{-0.3}$ \\
NGC5839 & $0.53^{+0.07}_{-0.07}$ & $1.79^{+0.09}_{-0.11}$ & $34^{+9}_{-8}$ & N & $...$ & $...$ & $...$ & $...$ & $0.85$ & $1.0^{+0.3}_{-0.2}$ \\
NGC5846 & $4.97^{+0.05}_{-0.06}$ & $1.69^{+0.06}_{-0.04}$ & $3198^{+139}_{-73}$ & Y & $3.60^{+0.26}_{-0.29}$ & $2.04^{+0.10}_{-0.10}$ & $0.75$ & $1.01$ & $...$ & $12.1^{+0.5}_{-0.3}$ \\
NGC5866 & $1.36^{+0.92}_{-0.44}$ & $7.33^{+0.45}_{-0.80}$ & $383^{+89}_{-81}$ & Y & $2.64^{+1.46}_{-0.83}$ & $0.59^{+0.38}_{-0.21}$ & $0.76$ & $1.05$ & $...$ & $2.2^{+0.5}_{-0.5}$ \\
PGC058114 & $0.65^{+1.03}_{-0.48}$ & $2.29^{+3.60}_{-1.61}$ & $20^{+29}_{-17}$ & N & $...$ & $...$ & $...$ & $...$ & $0.65$ & $1.0^{+1.4}_{-0.8}$ \\
NGC6548 & $0.26^{+0.54}_{-0.16}$ & $5.84^{+1.60}_{-2.59}$ & $9^{+23}_{-7}$ & N & $...$ & $...$ & $...$ & $...$ & $nan$ & $0.1^{+0.3}_{-0.1}$ \\
NGC7280 & $0.29^{+1.25}_{-0.18}$ & $2.09^{+2.92}_{-1.44}$ & $8^{+29}_{-6}$ & N & $...$ & $...$ & $...$ & $...$ & $nan$ & $0.2^{+0.6}_{-0.1}$ \\
NGC7332 & $0.60^{+0.18}_{-0.18}$ & $0.98^{+1.47}_{-0.62}$ & $117^{+47}_{-32}$ & Y & $1.23^{+1.15}_{-0.43}$ & $0.72^{+0.19}_{-0.30}$ & $0.77$ & $0.96$ & $...$ & $0.9^{+0.4}_{-0.2}$ \\
NGC7457 & $1.07^{+0.49}_{-0.27}$ & $4.12^{+2.69}_{-2.22}$ & $164^{+59}_{-50}$ & N & $...$ & $...$ & $...$ & $...$ & $0.80$ & $4.4^{+1.6}_{-1.4}$ \\
NGC7454 & $1.23^{+0.84}_{-0.41}$ & $5.86^{+1.52}_{-2.62}$ & $96^{+46}_{-36}$ & N & $...$ & $...$ & $...$ & $...$ & $0.97$ & $2.2^{+1.1}_{-0.8}$ \\
\enddata
\end{deluxetable*}

The effective radii of GC systems ($R_{e,gc}$) vary from sub-arcminutes to approximately 16 arcminutes. 
For instance, NGC4649 has the largest $R_{e,gc}$, and VCC1539 has the smallest $R_{e,gc}$, on an arcminute scale. 
S\'ersic $n$ values also vary within our fitting range, with most targets having S\'ersic $n$ values ranging from approximately 0.5 to 4. However, many program objects show substantial errors in their measured S\'ersic $n$ values due to a small number of GCs. The total number of GCs ranges from less than ten, for the faint dwarfs, to ten thousand or more for the brightest giants. NGC4486 has the largest number of GCs, while VCC1993 has the smallest number of GCs.

For each program galaxy, results are shown in a series of four figures. These are:
(1) A two-dimensional number density map of GC candidates. A mask map, which we used for S\'ersic fitting, is overlaid on the number density map.; (2) A two-dimensional and marginalized posterior probability density function. $R_{e,gc}$, S\'ersic $n$, and constant background are shown; (3) A one-dimensional radial GC number density profile with fitted model. In each case, we show the fitted S\'ersic function, the best-fit constant background, and their sum; and (4) A $(g-i)$ color distribution of GC candidates within $2.5R_{e,gc}$. We show the two fitted Gaussian functions and their peaks if it is bimodal. If the color distribution is unimodal, then we show a location of median color. 
These four figures for all targets are presented in Figure Set 10. in Appendix.
The exceptions are Figures ~\ref{fig:n0524},\ref{fig:n0821} which serve as representative examples for our program galaxies.

We make notes for individual galaxies in the Appendix, but here we point out some notable features or peculiarities for our sample galaxies:

\begin{itemize} 

\item {\bf Peculiar spatial distributions:} NGC1023, NGC4442, and NGC4608 display elongated spatial distributions of GCs. The GC number density peaks in NGC2685, VCC200, IC3328, VCC1512, VCC1833, UGC7854, VCC1993, and PGC058114 are offset from the galaxy centers. NGC3098, on the other hand, appears to show a lopsided distribution of GCs.

\item {\bf Exceptional color distributions:} NGC4564 contains a notably large population of red GCs.  NGC4694 also exhibits a large fraction of red GCs, although in this case, the results might be influenced by the existence of a large population of green GCs. NGC7454 stands out as having an unimodal GC population containing relatively red GCs.

\item {\bf Sparse GC Systems:} NGC6548 and NGC7280 contain almost no GCs within $2_{Re,GC}$, which may indicate that our fitting results are unreliable for these galaxies.

\end{itemize}

\begin{figure*}
\gridline{\fig{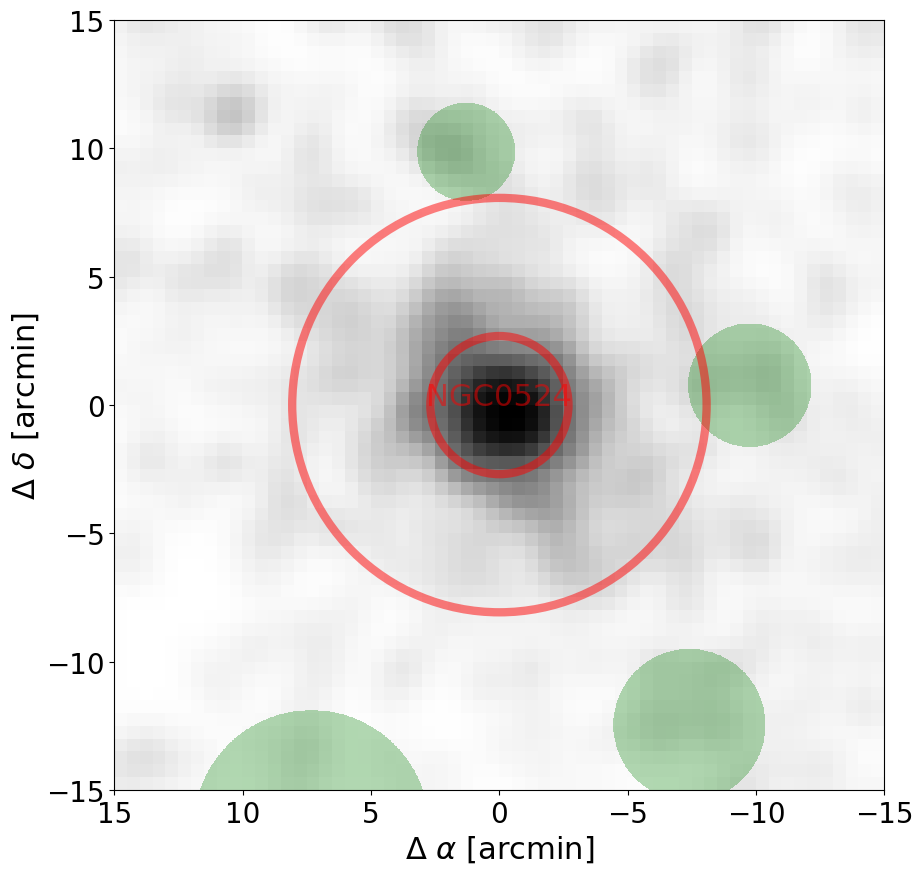}{0.48\textwidth}{(a)}
          \fig{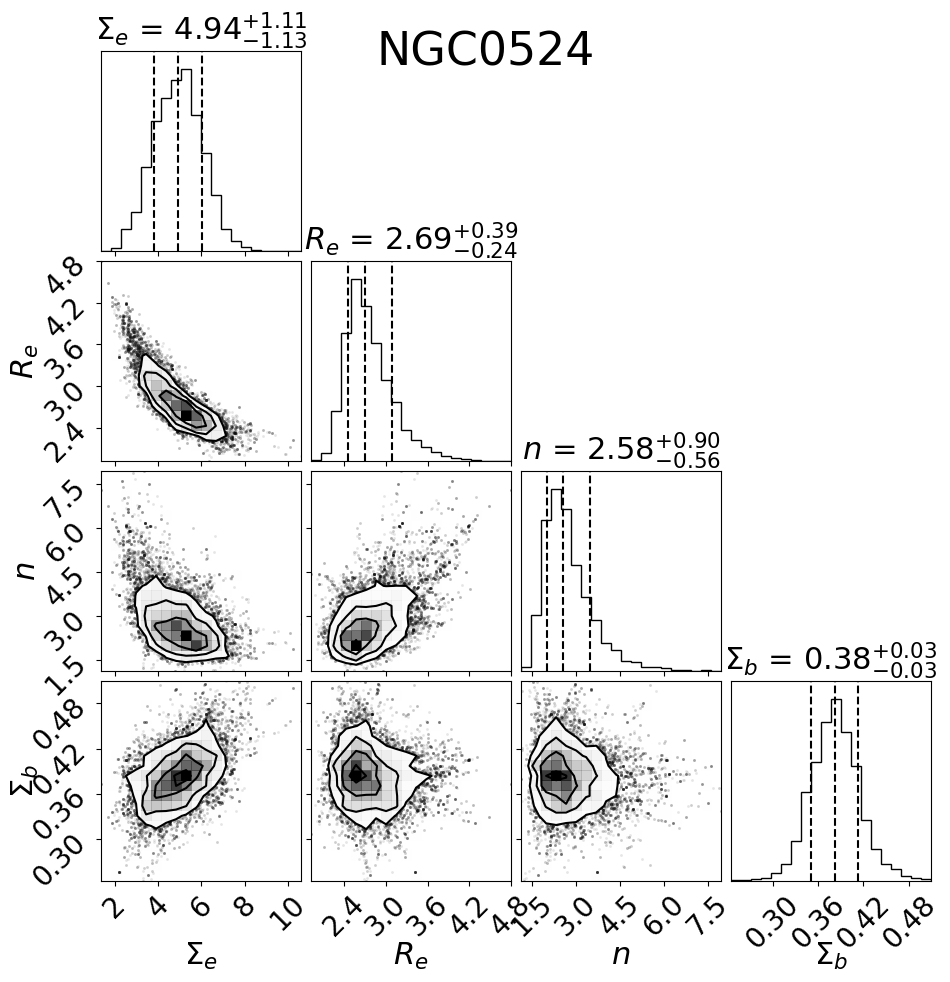}{0.48\textwidth}{(b)}
          }
\gridline{\fig{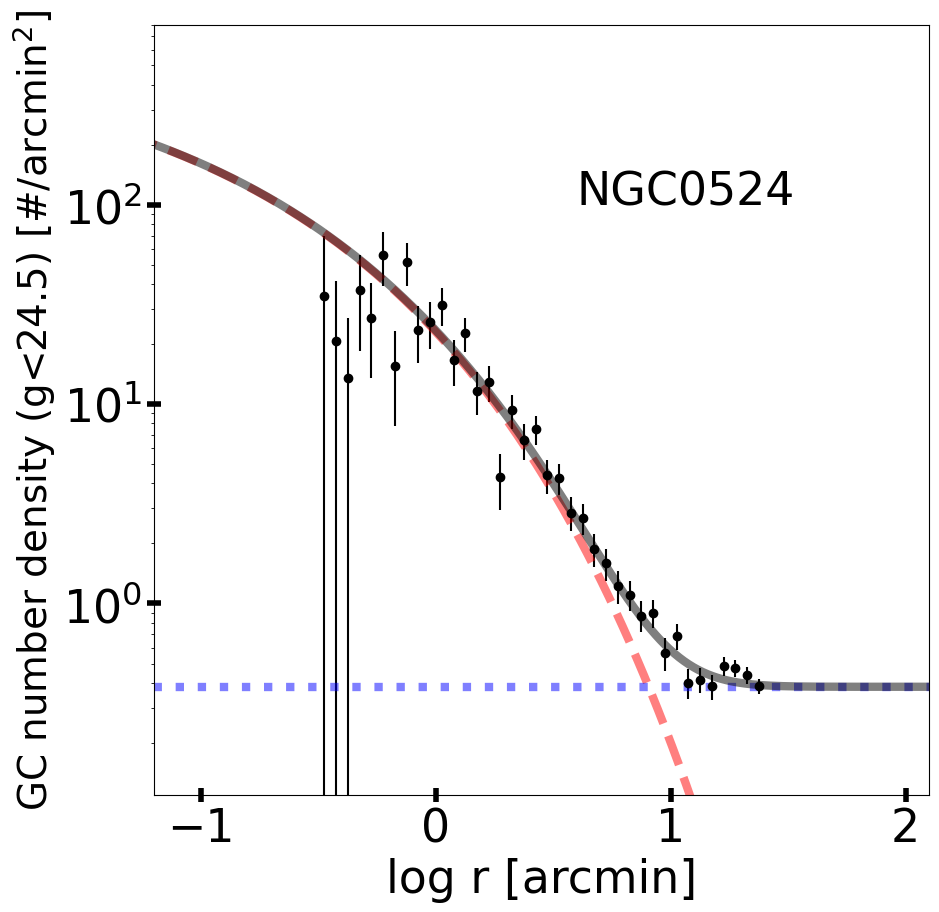}{0.48\textwidth}{(c)}
          \fig{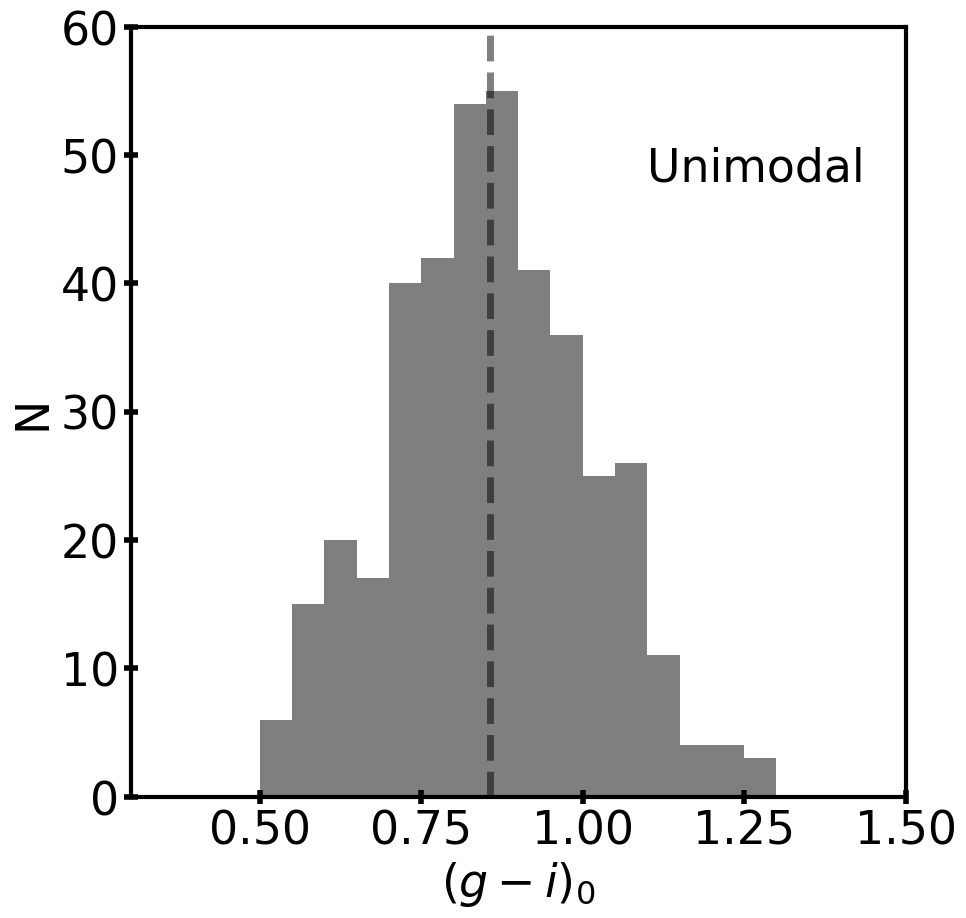}{0.48\textwidth}{(d)}
          }
\caption{(a) Number density map of GC candidates in the NGC524 region. The colorbar is on logarithmic scale. Red circles represent $1R_{e, GC}$ and $3R_e{,GC}$, respectively. Green shaded areas show the masked regions; (b) Two-dimensional and marginalized posterior probability density functions for the number density at the effective radius ($\Sigma_e$), the effective radius of the GC system ($R_{e}$), S\'ersic index ($n$), and constant background ($\Sigma_b$). The vertical lines represent 15th, 50th, and 84th quartiles from left to right; (c) One dimensional radial number density profile of GC candidates. Logarithmic bins are used. The black solid, red dashed, and blue dotted curves show the total function, S\'ersic function, and constant background, respectively; (d) $(g-i)_0$ color distribution of GCs within $2R_{e,gc}$ of NGC524. This distribution is categorized as a unimodal distribution by GMM. The black dashed line shows the median GC color.
\label{fig:n0524}}
\end{figure*}

\begin{figure*}
\gridline{\fig{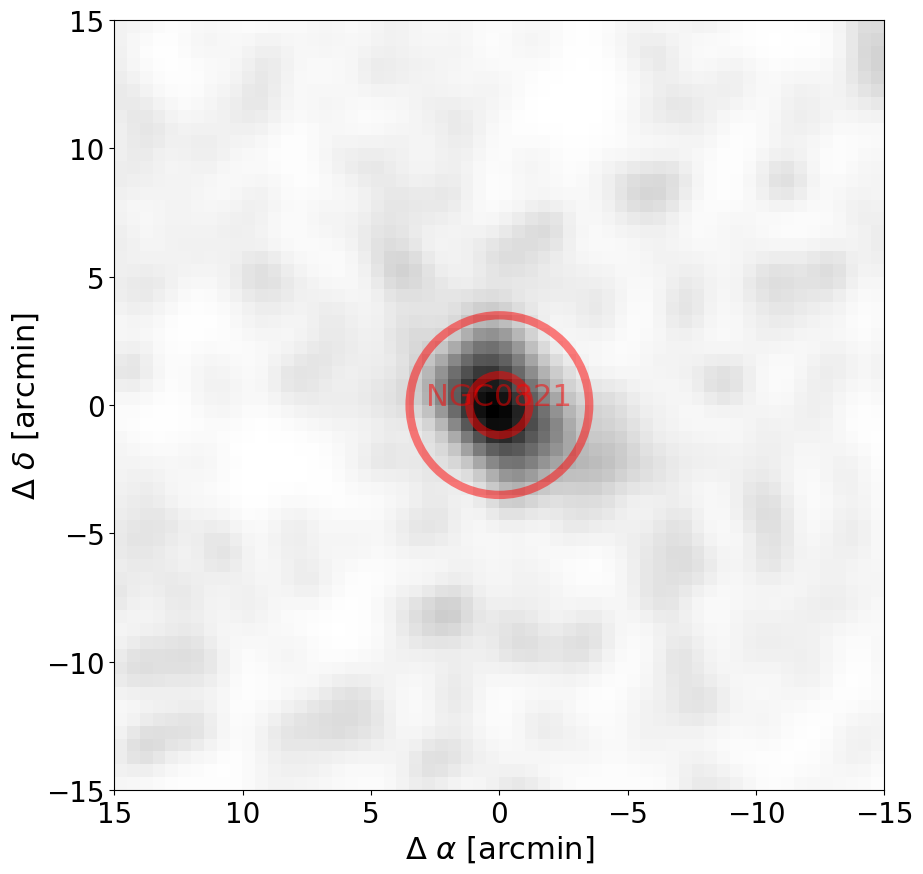}{0.48\textwidth}{(a)}
          \fig{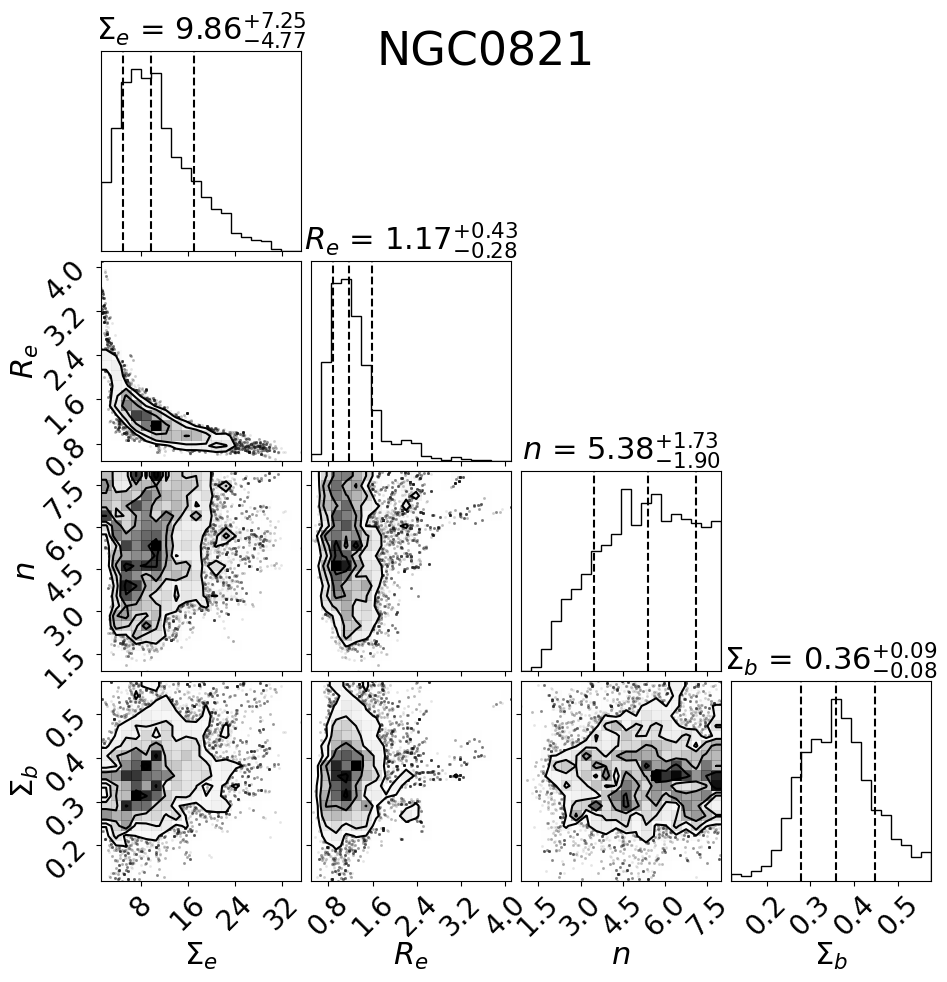}{0.48\textwidth}{(b)}
          }
\gridline{\fig{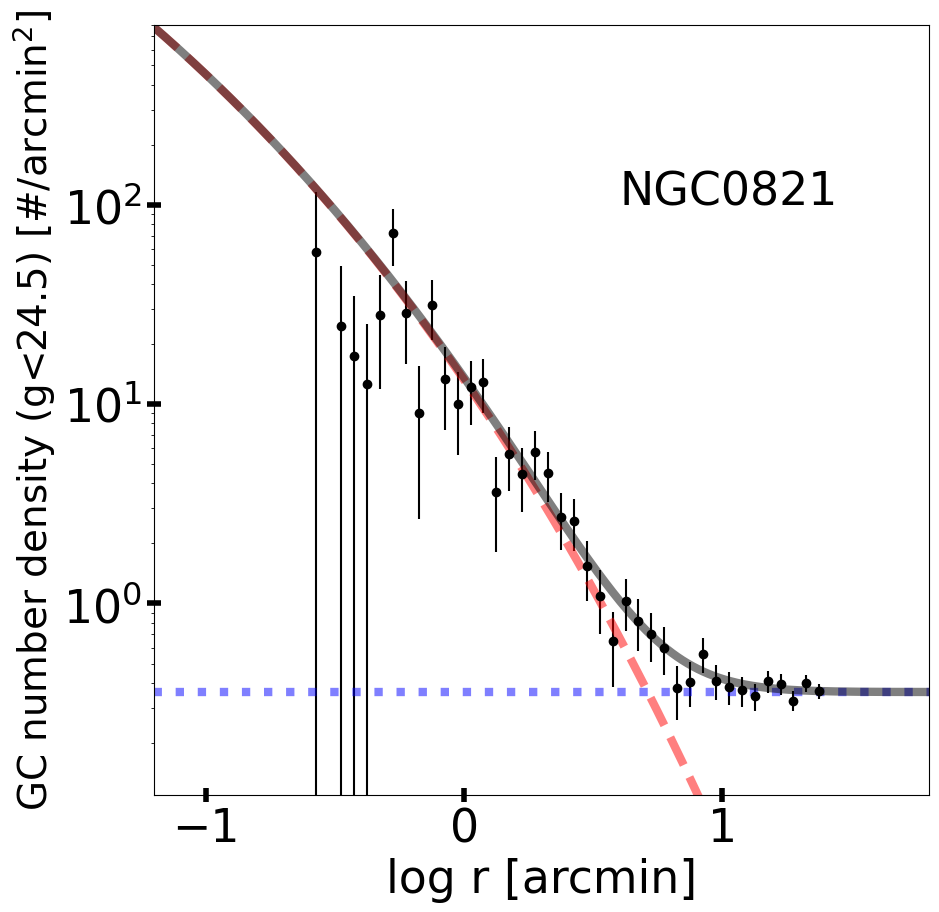}{0.48\textwidth}{(c)}
          \fig{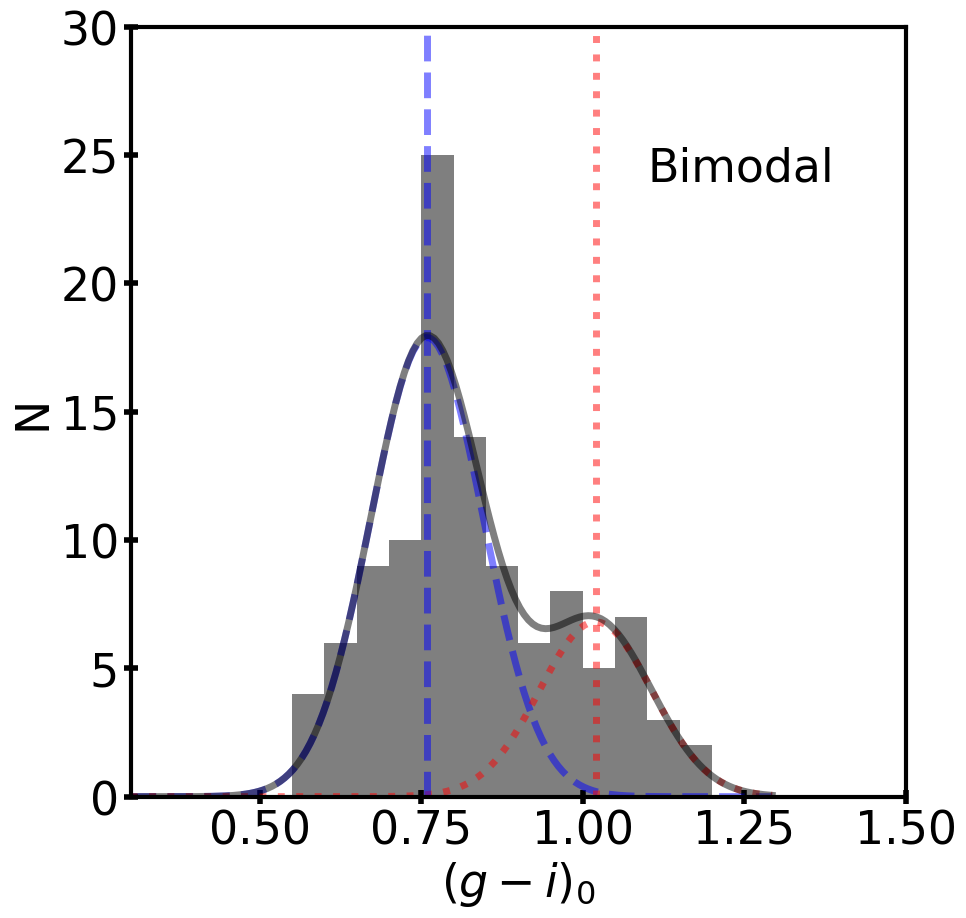}{0.48\textwidth}{(d)}
          }
\caption{(a) Number density map of GC candidates in the NGC821 region. The colorbar is on logarithmic scale. Red circles represent $1R_{e, GC}$ and $3R_e{,GC}$, respectively. Green shaded areas show the masked regions, but there is no masked region in NGC821; (b) Two-dimensional and marginalized posterior probability density functions for the number density at the effective radius ($\Sigma_e$), the effective radius of the GC system ($R_{e}$), S\'ersic index ($n$), and constant background ($\Sigma_b$). The vertical lines represent 15th, 50th, and 84th quartiles from left to right; (c) One dimensional radial number density profile of GC candidates in NGC821 region. The logarithmic bins are used. The black solid, red dashed, and blue dotted lines show the total function, S\'ersic function, and constant background, respectively; (d) $(g-i)_0$ color distribution of GCs within $2R_{e,gc}$ in NGC821. This distribution is categorized as a bimodal distribution by GMM. The blue dashed, red dotted, and black solid curves show fitted Gaussian functions of blue, red, and combined GCs, respectively. The vertical blue dashed and red dotted curves represent peak values of blue and red GC populations, respectively.
\label{fig:n0821}}
\end{figure*}

\begin{figure}
\epsscale{1.1}
\plotone{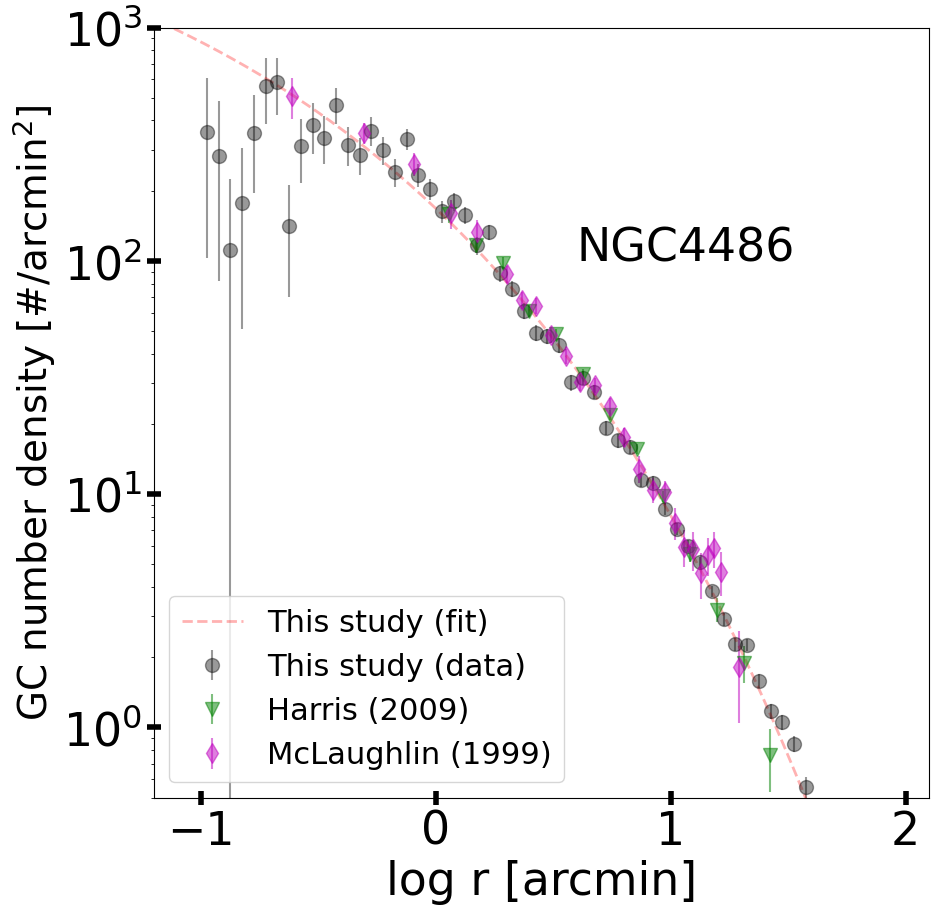}
\caption{Comparison of the GC number density profile in NGC4486 with the literature. X-axis is the radial distance from the center of NGC4486, and Y-axis is the GC number density. Gray circles and red dashed line show data points and fitted results from this study. For comparison with previous studies, the background is subtracted and the effect of the magnitude limit is corrected with the GCLF for GC number density. Magenta diamond and green inverted triangle represent results from \citet{1999AJ....117.2398M} and \citet{2009ApJ...703..939H}, respectively.
\label{comp_n4486}}
\end{figure}

\begin{figure}
\epsscale{1.1}
\plotone{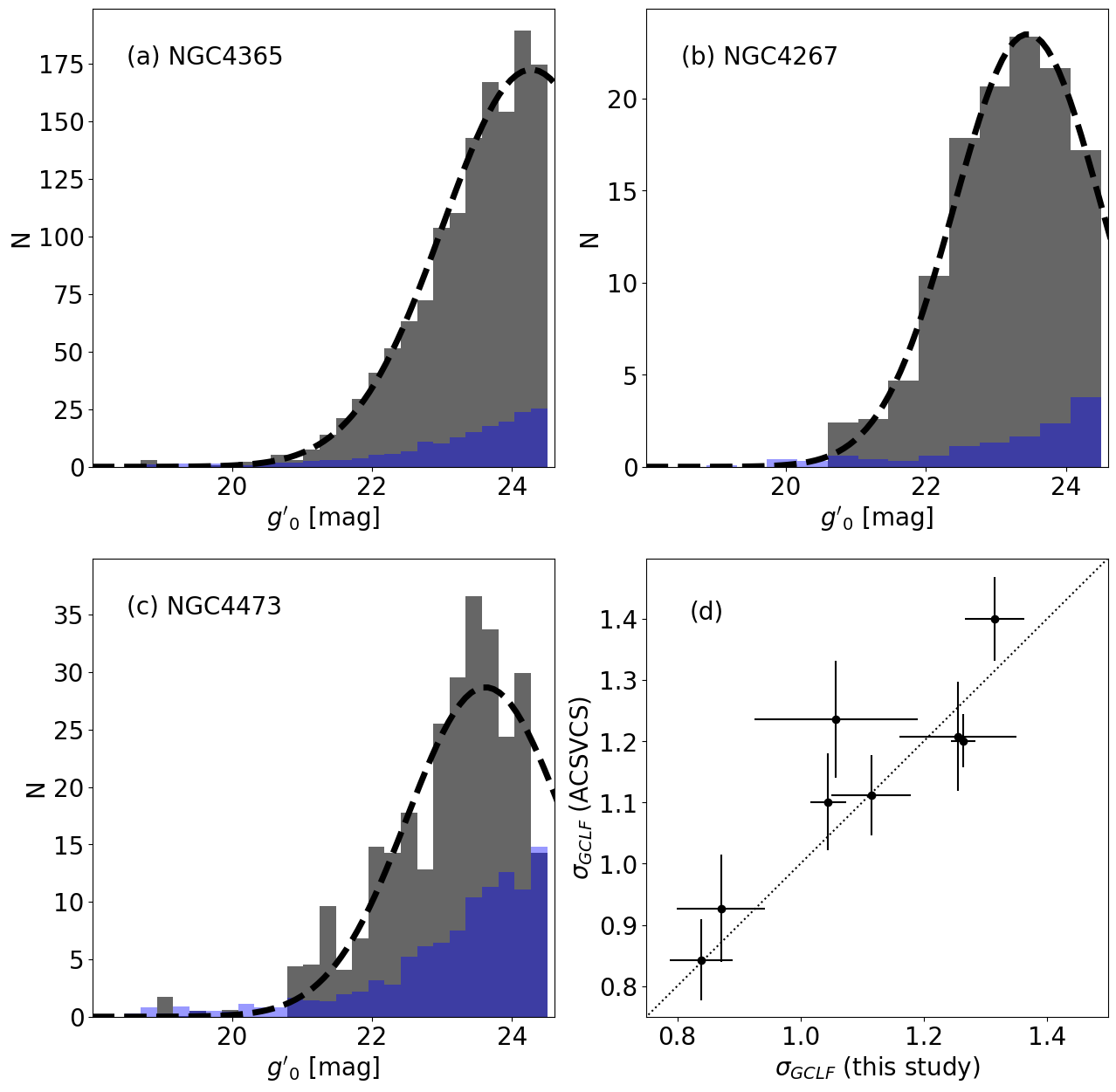}
\caption{GC luminosity functions and their fitting results, with the fixed peak magnitude of the Gaussian function. (a) GCLF of NGC 4365. The gray histogram shows the background-subtracted GCLF, and the blue histogram represents the background. GCs are within 2$R_{e,GCS}$, and the background area is chosen between 5$R_{e,GCS}$ and 8$R_{e,GCS}$. The dashed line represents the fitted Gaussian function with a fixed peak magnitude. (b) GCLF of NGC 4267, with notations the same as in panel (a). (c) GCLF of NGC 4473, with notations also the same as in panel (a). (d) Comparison of GCLF widths from this study and \citet{2010ApJ...717..603V}. The X and Y axes show $\sigma_{GCLF}$ from this study and \citet{2010ApJ...717..603V}, respectively. The dotted line indicates a one-to-one correspondence.
\label{gclf}}
\end{figure}

We compare our GC number density profile of NGC4486 with those from the literature to check for consistency (Figure~\ref{comp_n4486}). Each study has different magnitude limits in different filter systems, so all data points in Figure~\ref{comp_n4486} are background subtracted and corrected to full GCLF. All three data sets (this study, \citet{1999AJ....117.2398M} and \citet{2009ApJ...703..939H}) show consistent results.
To assess the reliability of using the GCLF parameters from \citet{2010ApJ...717..603V}, we also estimated the sigma of the GCLFs using our data and compared these values with those reported by \citet{2010ApJ...717..603V} (Figure \ref{gclf}). The sigma values from our study are consistent with those of \citet{2010ApJ...717..603V}, supporting our decision to use their GCLF parameters to estimate the total number of GCs.

\begin{figure}
\epsscale{1.1}
\plotone{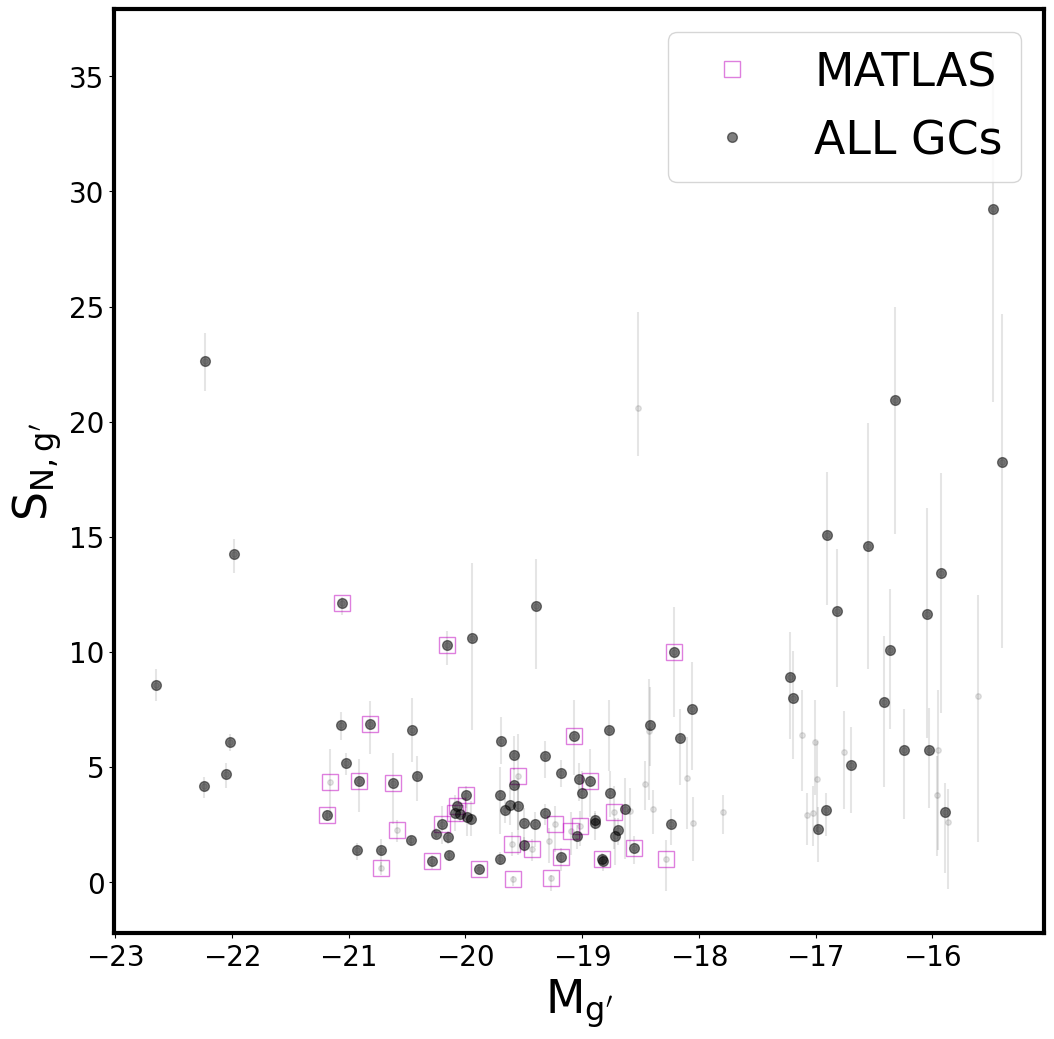}
\caption{The GC specific frequencies ($S_{N,g'}$) are plotted against the absolute $g'$-band magnitudes of their host galaxies. Gray filled circles show GC systems from this study; well-fitted systems (error of $R_{e,gc} < 50\%$) are shown with large heavy symbols, and MATLAS galaxies are highlighted.
\label{fg_sn}}
\end{figure}

We also estimated GC specific frequency of our samples. Traditionally, GC specific frequency is calculated with the $V$-band absolute magnitude of the host galaxy, but we estimated it with $g'$-band absolute magnitude of the host galaxy. We calculate GC specific frequency ($S_{N,g'}$) with a following equation:
\begin{equation}
    S_{N,g'} = N_{GC} \times 10^{0.4(M_{g'}+15)}
\end{equation}
Figure~\ref{fg_sn} shows relations between host galaxy magnitudes and $S_{N,g'}$. It shows a typical U-shape, and M87 has the highest $S_{N,g'}$ among massive galaxies. We marked MATLAS samples, and they mostly have low $S_{N,g'}$ except for three galaxies (NGC4278, NGC4283, and NGC5846). 

\subsection{Comparison of the total number of GCs with the ACSVCS} \label{compacsvcs}

\begin{figure*}
\epsscale{1.1}
\plotone{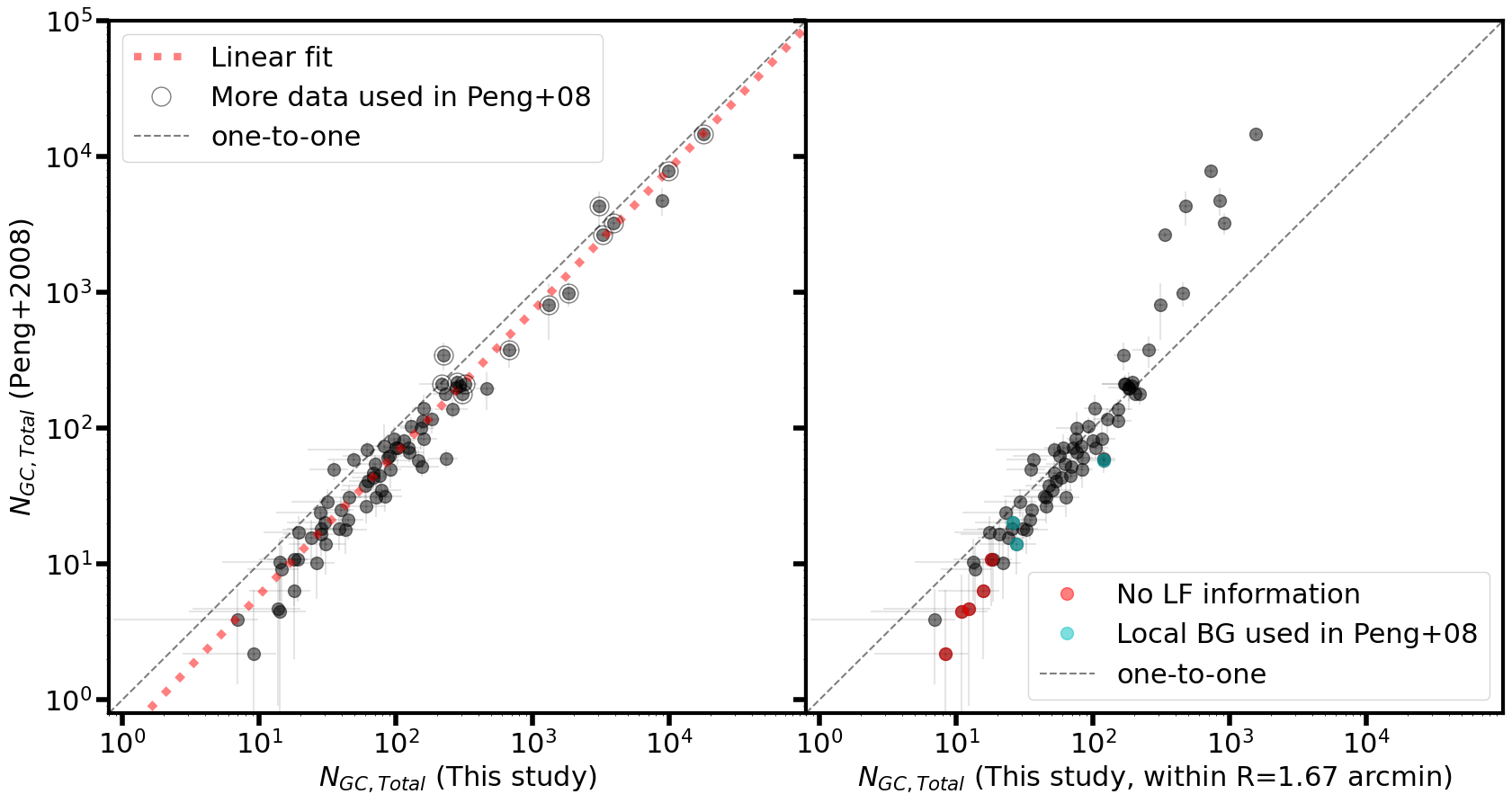}
\caption{Comparison of $N_{GC, total}$ from the ACSVCS results \citep{2008ApJ...681..197P} with this study. (left panel): Direct comparison of the total number of GCs from between this study and ACSVCS. The black dashed line and the red dotted line show the one-to-one relation and the linear fit, respectively. \citet{2008ApJ...681..197P} used additional data to fit GC density profiles for several galaxies because of large spatial distribution of GCs, and we mark them with large open circles.
(right panel): Similar to the left panel, but total numbers of GCs in this study are limited within $1.67\arcmin$ similar to the ACS/WFC field of view. The dashed line shows the one-to-one relation. \citet{2008ApJ...681..197P} used additional background observations to estimate the number of GCs, but they used local background (in an ACS/WFC field) for several galaxies, so these galaxies are marked with cyan filled circles. This study uses Gaussian functions of the ACSVCS GCLF to correct for the magnitude limit, but several ACSVCS galaxies do not have a Gaussian form of the GCLF, so we have marked them with red filled circles. 
\label{acsvcs}}
\end{figure*}

Seventy-five targets in this study overlap with those in the ACSVCS, providing the total number of GCs. We compare these total numbers with those from \citet{2008ApJ...681..197P} in Figure~\ref{acsvcs}. The direct comparison shows that this study's total number of GCs is slightly larger than that in the ACSVCS. The agreement is better for larger numbers, as the ACSVCS includes additional data outside the ACS/WFC field of view (FoV) for large galaxies (marked with large open circles in the left panel of Figure~\ref{acsvcs}). While we might expect good agreement for medium and low-number GCs, the differences in total GC numbers between this study and the ACSVCS increase as the total number of GCs decreases. 

To investigate this trend, we compare the total number of GCs in the ACSVCS with the results from this study but limit it to GCs within $1.67\arcmin$ from the galaxy center to match the ACS/WFC FoV, as shown in the right panel of Figure~\ref{acsvcs}. Galaxies with about one hundred GCs have consistent results between the ACSVCS and this study, suggesting that the number of GCs in intermediate-luminosity galaxies with $N_{GC,Total} \approx 100$ may be underestimated in the ACSVCS due to the limit of the ACS FoV. However, at the lower end of the number scale, the discrepancy between the two studies persists even with similar area coverage, indicating that the different coverages of both studies does not account for this difference. 

We investigated additional factors that could contribute to these differences. In the case of several low-mass galaxies, GC numbers were estimated under specific conditions. In the ACSVCS, additional background fields were used to account for background contamination when estimating GC numbers. However, for several low-mass galaxies, a local background was employed instead. These galaxies are marked with cyan circles in the right panel of Figure~\ref{acsvcs}. In our GC number estimation, we used the fitted GCLFs from the ACSVCS for GCLF correction. However, there were a couple of galaxies in the ACSVCS for which GCLF information was unavailable. In these cases, we used the relation between GCLF and the host galaxy luminosity, marked with red in the right panel of Figure~\ref{acsvcs}. About half of the low-GC-number galaxies had special conditions for GC number estimation in the ACSVCS. This suggests that the discrepancy in the number of low-GC galaxies may be due to these special conditions.

There remain, however, several low-mass galaxies with discrepancies in GC numbers even when particular conditions are not involved. It is challenging to verify the cause of these discrepancies. We must delve into the differences in GC selection methods between the two studies to gain more insight. In our study, we used color-color and size information for GC selection. Additionally, we extended the maximum size limit for Virgo galaxies. The GC selection process used in the ACSVCS is more complicated than this study. They define the GC probability with color, size, and shape information. Based on this GC probability, they could identify clean GCs within specific color ranges and round, compact, but resolved sources. However, recent studies indicate that GCs can exhibit larger sizes and elongated shapes (e.g., \citealp{2014ApJ...794..103D,2018ApJ...856L..30V}), potentially leading the ACSVCS survey to exclude some genuine GCs. This study included large GCs and did not consider ellipticity in GC selection, which should result in more complete GC samples than the ACSVCS. This effect of unusual GCs may be particularly significant in galaxies with fewer GCs. 

\subsection{Comparison of the effective radii with the literature}
\begin{figure}
\epsscale{1.1}
\plotone{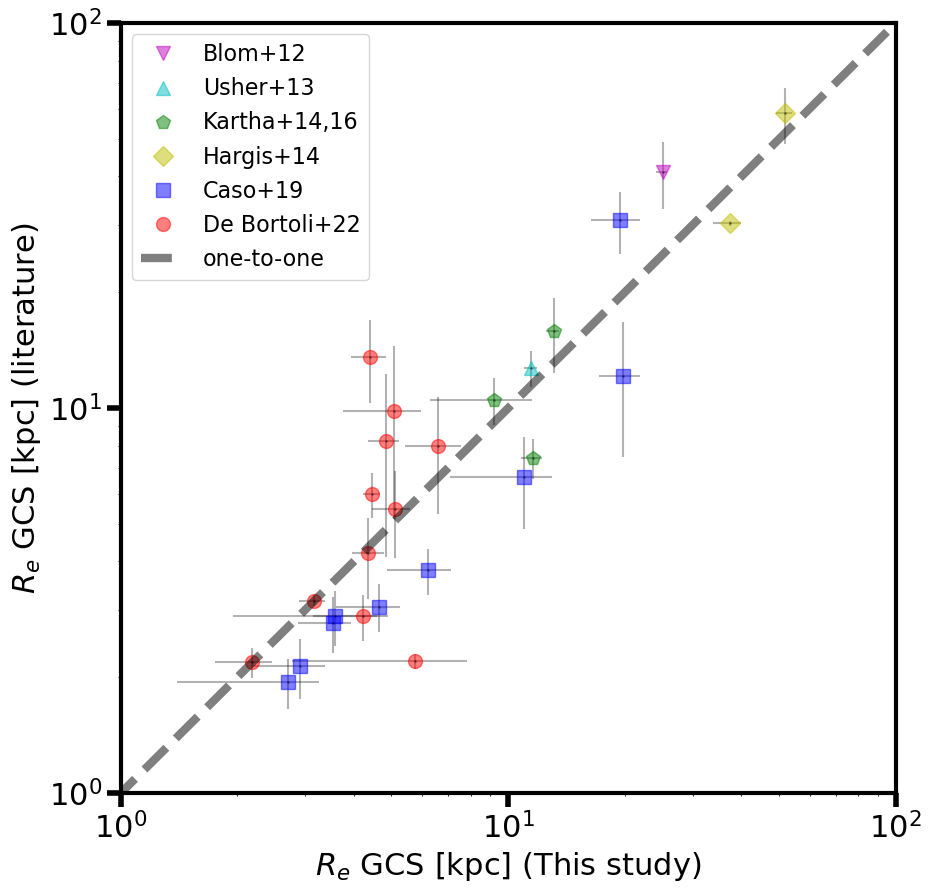}
\caption{Comparison of $R_{e,gc}$ from literature with this study. The downward triangle shows NGC4365 \citep{2012MNRAS.420...37B}; the triangle shows NGC4278 \citep{2013MNRAS.436.1172U}, and pentagons show three galaxies from the SLUGGS survey \citep{2014MNRAS.437..273K,2016MNRAS.458..105K}, diamonds, NGC4472 and NGC4406 \citep{2014ApJ...796...62H}, squares \citep{2019MNRAS.488.4504C}, and circles \citep{2022MNRAS.510.5725D}
\label{recomplit}}
\end{figure}

Previous studies have reported effective radii for the GC systems in some of our program galaxies. Figure~\ref{recomplit} compares the effective radii for GC systems from the literature with those measured in this study. Generally speaking, our measurements are consistent with those in the literature, albeit with some scatter. There are, however, slight differences between some individual literature measurements. In literature studies examining multiple galaxies, the findings of \citet{2014MNRAS.437..273K,2016MNRAS.458..105K} show two cases consistent with our results and one notable discrepancy. The outlier is NGC 3608, which has a nearby companion galaxy, NGC 3607, with a GC system approximately twice as large as that of NGC 3608. \citet{2014MNRAS.437..273K,2016MNRAS.458..105K} separated the GCs of the two galaxies based on their spatial locations and fit their distributions independently. However, as the GCs from both galaxies overlap in spatial extent, a simultaneous fitting approach we employed in this study is necessary for more accurate results. \citet{2022MNRAS.510.5725D} results also exhibit considerable divergence from ours, primarily due to three galaxies with the largest GC systems in their study and one with a notably smaller GC system than observed in our analysis. The largest GC system reported by \citet{2022MNRAS.510.5725D} is for NGC 4435, which has a nearby companion that may contribute additional GCs, thus increasing the size of the GC system. The two other large GC systems in \citet{2022MNRAS.510.5725D} show substantial uncertainties in size estimation, leading us to consider our estimates more reliable. Additionally, the smallest GC system reported by \citet{2022MNRAS.510.5725D} may be influenced by the limited spatial coverage of their data, an issue similarly noted in comparison with results from \citet{2019MNRAS.488.4504C}.

\section{Summary}

We investigate the spatial distribution of GCs belonging to 118 early-type galaxies using imaging data from the NGVS, and the MATLAS, supplemented by the ACS Virgo Cluster Survey for central galaxy regions, when available. Our program, along with a companion paper, aims to understand the connection between galaxy evolution and GC system size, focusing on the effective radii of GC systems and their correlation with various galaxy properties \citep{2024ApJ...966..168L}. 

Photometry is performed on model-subtracted galaxy images to detect GC candidates. Aperture magnitudes are used to estimate source fluxes, with corrections made for fixed aperture size limitations using the largest available aperture. We select GC candidates based on both size/concentration and two-color information, considering both point-like and slightly extended sources.

Completeness tests are carried out by injecting approximately 200,000 artificial stars into each field. The recovery rate for the brighter than $m_g =24.5$ mag is almost $100\% $ across the overall field except for the central regions of bright galaxies.

Spatial distributions of GCs are modeled using modified two-dimensional S\'ersic profiles. Our fitting procedure employs the MCMC method, explicitly utilizing the emcee code in Python. In cases where neighboring galaxies harbor a significant GC population, simultaneous fitting with two S\'ersic functions is applied. We estimate the total number of GCs by integrating the S\'ersic profiles and correcting for the GCLF.

Furthermore, GCs are categorized into sub-samples based on their colors using the GMM. More than half of the sample (68) show bimodality in GC color distribution. Spatial distributions of these sub-samples are also subjected to fitting.

This study represents the largest and most homogeneous sample to date for studying the spatial distributions of GC systems. A companion paper \citep{2024ApJ...966..168L} provides the scientific outcomes from the data and catalogues described in this paper. In a future paper, we plan to extend the analysis to include dwarf galaxies outside galaxy cluster environments, explicitly focusing on investigating the GC spatial distributions for dwarf galaxies using MATLAS data. Euclid will soon provide a large database as well.

\clearpage
\begin{acknowledgments}
S.L. acknowledges the support from the Sejong Science Fellowship Program by the National Research Foundation of Korea (NRF) grant funded by the Korea government (MSIT) (No. NRF-2021R1C1C2006790).
CL acknowledges support from the National Natural Science Foundation of China (NSFC, Grant No. 12173025, 11833005, 11933003), 111 project (No. B20019), and Key Laboratory for Particle Physics, Astrophysics and Cosmology, Ministry of Education.
C.S. acknowledges support from ANID/CONICYT through FONDECYT Postdoctoral Fellowship Project No. 3200959.
AL and PAD acknowledge support from Agence Nationale de la Recherche, France, under project ANR-19-CE31-0022.
O.M. is grateful to the Swiss National Science Foundation for financial support under the grant number PZ00P2\_202104. 

IRAF was distributed by the National Optical Astronomy Observatory, which was managed by the Association of Universities for Research in Astronomy (AURA) under a cooperative agreement with the National Science Foundation.
Based on observations obtained with MegaPrime/MegaCam, a joint project of CFHT and CEA/DAPNIA, at the Canada-France-Hawaii Telescope (CFHT) which is operated by the National Research Council (NRC) of Canada, the Institut National des Science de l'Univers of the Centre National de la Recherche Scientifique (CNRS) of France, and the University of Hawaii. The observations at the Canada-France-Hawaii Telescope were performed with care and respect from the summit of Maunakea which is a significant cultural and historic site.

This research is based on observations made with the NASA/ESA Hubble Space Telescope obtained from the Space Telescope Science Institute, which is operated by the Association of Universities for Research in Astronomy, Inc., under NASA contract NAS 5–26555. These observations are associated with program GO-9401.
This research was supported by the International Space Science Institute (ISSI) in Bern, through ISSI International Team project \#534 (Space Observations of Dwarf Galaxies from Deep Large Scale Surveys: The MATLAS Experience). 

Some of the data presented in this paper were obtained from the Mikulski Archive for Space Telescopes (MAST) at the Space Telescope Science Institute. The specific observations analyzed can be accessed via \dataset[DOI: 10.17909/yc2t-rr81]{https://doi.org/10.17909/yc2t-rr81}.

\end{acknowledgments}

%

\facilities{HST/ACS/, CFHT}


\software{astropy \citep{astropy:2013,astropy:2018,astropy:2022},  
          Source Extractor \citep{1996A&AS..117..393B}
          }



\appendix
\section{Notes for all samples}

\begin{itemize} 

\item NGC524: The GC spatial distribution is well fitted by the S\'ersic function. Interestingly, it shows a unimodal GC color distribution even though it is a fairly massive galaxy. 
The central region of this galaxy is studied with HST \citep{2001AJ....122.1251K} and the GC color distribution in the HST study is similar to that of this study. It is notable that a previous spectroscopic study has demonstrated that metal-poor globular clusters (GCs) exhibit distinct kinematics when compared to metal-rich GCs \citep{2004MNRAS.347.1150B}. To confirm the existence of GC subpopulations, further spectroscopic observations of GCs are necessary.

\item NGC821: The GC spatial distribution is well fitted by the S\'ersic function with a high S\'ersic index. The number of GCs ($N_{GC}=764^{+315}_{-230}$) in this study is lager than twice of those in the literature ($N_{GC}=395\pm94$, \citealp{2001AJ....121.2950K}; $N_{GC}=320\pm54$, \citealp{2008MNRAS.385..361S}). This discrepancy may be attributed to differences in the coverage and depth of observation. The radial number density profile of GCs in the literature reaches a background level at approximately three arcmin, whereas our findings indicate an excess of GCs up to 10 arcmin.

\item NGC936: GCs are more centrally concentrated than other galaxies.

\item NGC1023: The GC spatial distribution is elongated. The GC system in this galaxy has been studied in several literature \citep{2000AJ....120.2938L,2012AJ....144..103Y,2014MNRAS.437..273K,2022A&A...664A.129D}. The number of GCs in this study is slightly smaller than that reported in the literature. The reason for this discrepancy is not immediately apparent. However, it is possible that it may be due to differences in the fitting of radial profiles. 

\item NGC2592: There is a background galaxy (NGC2594, 35.1Mpc) to the southwest of NGC2592. We masked out this region instead of fitting it simultaneously because of its different distance.

\item NGC2685: The GC number density peak is offset from the galaxy center.

\item NGC2768: The S\'ersic function with a high S\'ersic index fits the GC spatial distribution well. This galaxy is studied in \citet{2014MNRAS.437..273K}, and the effective radius of the GC system is consistent, but the number of GCs in this study is $\sim40\%$ larger than that in \citet{2014MNRAS.437..273K}.

\item NGC2778: The GC spatial distribution has a long tail to the north. The GC color distribution is determined to be bimodal by statistical testing, but the red population is very small.

\item NGC2950: The blue and red GC populations are comparable.

\item NGC3098: The GC number density peak is offset from the galaxy center. The blue and red peaks of GCs are relatively bluer than those of other galaxies.

\item NGC3245: The GC color distribution is unimodal, but has a peak in blue and a tail to red.

\item NGC3379: It has a neighbor, NGC3384, so we fit two S\'ersic functions simultaneously. The fitted S\'ersic index is close to 8, which is our upper limit for the S\'ersic index. The blue and red GC populations are comparable. This galaxy is studied in \citet{2004AJ....127..302R}, and the total number of GCs is comparable in both studies.

\item NGC3384: It has a neighbor, NGC3384, so we fit two S\'ersic functions simultaneously. Interestingly, its fitted S\'ersic index is about 0.5, which is close to our lower limit for the S\'ersic index. The red GC population is larger than the blue GC population. \citet{2012AJ....144..164H} studied this galaxy and their result for the total number of GCs is consistent with this study.

\item NGC3457: There is no special feature in GC properties.

\item NGC3489: The GC color distribution is very clearly bimodal, and the color of red GCs is redder than that of other galaxies. \citet{2001AJ....122.1251K} studied the central region of this galaxy with the HST, but it is difficult to find distinguished red GCs in this literature.

\item NGC3599: There is no special feature in GC properties. The central region of this galaxy is studied by \citet{2001AJ....122.1251K} with the HST.

\item NGC3607: It has a neighbor, NGC3608, so we fit two S\'ersic functions simultaneously. The fitted S\'ersic index is close to 8, which is our upper limit for the S\'ersic index. \citet{2016MNRAS.458..105K} has studied this galaxy, and their $R_{e,gc}$ is consistent with this study, but their S\'ersic index is much smaller than us. GCs in the central region were studied in \citet{2001AJ....122.1251K}.

\item NGC3608: It has a neighbor, NGC3607, so we fit two S\'ersic functions simultaneously. \citet{2016MNRAS.458..105K} has studied this galaxy and their $R_{e,gc}$ is a little smaller than this study. GCs in the central region were studied in \citet{2001AJ....121.2950K}.

\item NGC3630: GCs are more centrally concentrated than other galaxies.

\item NGC3945: The GC number density peak is offset from the galaxy center.

\item IC3032: The GC number density peak is offset from the galaxy center. We have a larger total number of GCs than that in the ACSVCS \citep{2008ApJ...681..197P}. Please check the section \ref{compacsvcs}.

\item IC3065: The GC number density peak is offset from the galaxy center. The total number of GCs is about twice that in the ACSVCS \citep{2008ApJ...681..197P}. Please check the section \ref{compacsvcs}.

\item VCC200: The GC number density peak is offset from the galaxy center. The total number of GCs is consistent with that in the ACSVCS \citep{2008ApJ...681..197P}. 

\item IC3101: There is no special feature in GC properties. The total number of GCs is consistent with that in the ACSVCS \citep{2008ApJ...681..197P}. 

\item NGC4262: A small neighboring galaxy is in the field, so we masked that region. \citet{2024MNRAS.530.2907A} have studied GC properties using the same data as this study, but their results differ slightly from those in this study. This may be due to different analysis methods, including different magnitude limits and fitting methods (binned data versus individual data points).

\item NGC4267: The GC number density profile is flattened in the central region. It has a large population of red GCs. \citet{2022MNRAS.510.5725D} studied the $R_{e,gc}$ of this galaxy, but their $R_{e,gc}$ is slightly larger than in this study, which could be due to different data (HST only versus combined HST and CFHT).

\item NGC4278: It has a neighbor, NGC 4283, so we fit two S\'ersic functions simultaneously. Interestingly, $R_{e,BGC}$ is much smaller than $R_{e,gc}$, which could be due to the extremely large $R_{e,BGC}$ of its neighbor, NGC 4283. Although there is a high possibility of underestimating $R_{e,BGC}$ for NGC 4278 (and conversely, overestimating $R_{e,BGC}$ for NGC 4283), we keep these results to maintain consistency in estimating $R_{e,gc}$ for all samples.
\citet{2013MNRAS.436.1172U} have investigated the GC properties of this study. Their $R_{e,gc}$ is in agreement with this study.

\item NGC4283: It has a neighbor, NGC4278, so we fit two S\'ersic functions simultaneously. Please check the note for NGC4278.

\item UGC7436: GCs are more centrally concentrated than other galaxies. It is included in \citet{1996AJ....112..972D}, and they provided a smaller total number of GCs ($N_{GC} = 20\pm11$) that that of this study, but it is consistent with us within the margin of error. The total number of GCs in this study is larger than that in the ACSVCS ($N_{GC}=18.1\pm5.5$, \citealp{2008ApJ...681..197P}). Please check the section \ref{compacsvcs} for this discrepancy. 

\item VCC571: GCs are rarely detected. The total number of GCs is slightly larger than that in the ACSVCS \citep{2008ApJ...681..197P}. Please check the section \ref{compacsvcs}.

\item NGC4318: GC color distribution is bimodal even with a very small number of GCs, and this bimodality was also shown at \citet{2006ApJ...639...95P} .

\item NGC4339: There is no special feature in GC properties.

\item NGC4340: It has a neighbor, NGC4350, so we fit two S\'ersic functions simultaneously. The total number of GCs is slightly larger than that in the ACSVCS \citep{2008ApJ...681..197P}, but they are consistent within the margin of error. 

\item NGC4342: It has a neighbor, NGC4365, so we fit two S\'ersic functions simultaneously. The color distribution is bimodal, but blue and red peaks are very close. \citet{2014MNRAS.439.2420B} have studied GC properties of this galaxy, and their $R_{e,gc}$ and S\'ersic index $n$ values are consistent with this study.

\item NGC4350: It has a neighbor, NGC4340, so we fit two S\'ersic functions simultaneously. The total number of GCs is larger than that in the ACSVCS \citep{2008ApJ...681..197P}. Please check the section \ref{compacsvcs}.

\item NGC4352: There is a bright star nearby, so its region is masked. The total number of GCs is consistent with that in the ACSVCS \citep{2008ApJ...681..197P}. 

\item NGC4365: It has a neighbor, NGC4342, so we fit two S\'ersic functions simultaneously. The blue and red GC populations are comparable. The GC system of this galaxy have been studied in many literature (e.g. \citealp{1996AJ....112..954F,2001AJ....121.2950K,2002A&A...391..453P,2005AJ....129.2643B,2005ApJ...634L..41K}), and \citet{2012MNRAS.420...37B} provide the fitting results for the GC spatial distribution. Their $R_{e,gc}$ and $N_{GC}$ are much larger than those in this study. This discrepancy could be due to background estimation. They could not reach the edge of the GC spatial distribution in their observation, but we did. Additionally, they do not mention the effect of NGC 4342. If we fit the GC spatial distribution with a single S\'ersic profile while masking out the NGC 4342 area, we obtain a slightly larger $R_{e,gc}$ than from a two-S\'ersic fit. Therefore, we expect that the $R_{e,gc}$ in the literature was overestimated due to the effect of the neighboring galaxy.

\item NGC4371: There is no special feature in GC properties. \citet{2022MNRAS.510.5725D} provide $R_{e,gc}$, and it is much larger than that in this study. However, their $R_{e,gc}$ has a large error, so their $R_{e,gc}$ and that in this study are consistent within the margin of error.

\item NGC4374: It has a neighbor, NGC4406, so we fit two S\'ersic functions simultaneously. There are other galaxies nearby, so we masked them out aggressively. There are several published works that provide $N_{GC}$ of this galaxy (e.g. \citealp{2004A&A...415..499G,2008ApJ...681..197P,2020ApJ...900...45L}). However, previous literatures do not address the impact of NGC4406 due to limit of data, suggesting that our result could be the most reliable.

\item NGC4377: The blue and red GC populations are comparable, but it is not shown in the ACSVCS \citep{2006ApJ...639...95P}. \citet{2022MNRAS.510.5725D} shows $R_{e,gc}$ of this galaxy, and it is consistent with this study. 

\item NGC4379: There is no special feature in GC properties. The central region was investigated using HST data in \citet{2001AJ....122.1251K}, but the resulting $N_{GC}$ values are considerably smaller due to the limitations of the HST spatial coverage.

\item NGC4387: There are two big neighbors. Because both galaxies seem to affect equally, it is difficult to choose one of them for dual S\'ersic fit, so we masked them out. Total numbers of GCs from literatures \citep{2008ApJ...681..197P,2020ApJ...900...45L} are consistent with our result, and $R_{e,gc}$ from \citet{2019MNRAS.488.4504C} is also consistent with our result.

\item IC3328: There is no special feature in GC properties.

\item NGC4406: It has a neighbor, NGC4374, so we fit two S\'ersic functions simultaneously. There are other galaxies nearby, so we masked them out aggressively. $N_{GC}$ from this study is slightly larger than those in literature ($N_{GC}=2900\pm400$\citealp{2004AJ....127..302R,2020ApJ...900...45L}; $N_{GC}=2660\pm129$, \citealp{2008ApJ...681..197P}), and it may be due to the limited spatial coverage of previous studies. \citet{2014ApJ...796...62H} estimated $R_{e,gc}=5\arcmin.8\pm0\arcmin.1$ with S\'ersic function fitting, that is smaller than our result. They also provide an empirical $R_{e,gc} = 6\arcmin.4$, which is still slightly smaller than our result. This discrepancy may be due to differences in the approach to handling neighboring galaxies.

\item NGC4417: It has a neighbor, NGC4424, so we fit two S\'ersic functions simultaneously. \citet{2022MNRAS.510.5725D} estimated $R_{e,gc}$ of this galaxy, that is consistent with that in this study.

\item NGC4425: It has a neighbor, NGC4406, so we fit two S\'ersic functions simultaneously.

\item NGC4429: It has a bimodal GC color distribution, but there is a peak in the middle of the blue and red GCs.

\item NGC4434: There is a bright star nearby, so its region is masked. $N_{GC}$ and $R_{e,gc}$ from previous studies \citep{2008ApJ...681..197P,2019MNRAS.488.4504C} are smaller than those in this study. It may be due to the limited spatial coverage of previous studies.

\item NGC4435: The west side is close to NGC4406, and there is significant background contamination on the east side, so we aggressively masked it out. It has a neighbor, NGC4438, so we fit two S\'ersic functions simultaneously. It is interesting that $N_{GC}$ and $R_{e,gc}$ from previous studies \citep{2008ApJ...681..197P,2022MNRAS.510.5725D} are larger than those in this study. This discrepancy may be due to the different methods employed in dealing with the background and neighbor galaxies.

\item NGC4442: The GC spatial distribution is elongated. The blue and red GC populations are comparable, while this galaxy has no particular high red GC fraction in the ACSVCS \citep{2006ApJ...639...95P}. $N_{GC}$ from the ACSVCS \citep{2008ApJ...681..197P} is smaller than that in this study, and it could be due to the small areal coverage of the ACSVCS. $R_{e,gc}$ from the previous study ($R_{e,gc}=1\arcmin.8 \pm 0\arcmin.9$, \citealp{2022MNRAS.510.5725D}) is much larger than our result, but both are consistent within the margin of error.

\item IC3383: There is a neighbor dwarf galaxy at North, so we masked it out.

\item IC3381: There is no special feature in GC properties.

\item NGC4452: There is a neighboring dwarf galaxy and a bright star, so we masked them out.

\item NGC4458: We rigorously masked out due to three nearby bright neighbors. 
$N_{GC}$ in this study is a little larger than those in the previous studies \citep{2001AJ....121.2950K,2008ApJ...681..197P}, and it may be due to the limited spatial coverage of the previous studies. $R_{e,gc}$ in the previous study \citep{2019MNRAS.488.4504C} is consistent with that in this study.

\item NGC4459: The blue and red GC populations are comparable, while this galaxy has no particular high red GC fraction in the ACSVCS \citep{2006ApJ...639...95P}. $N_{GC}$. $N_{GC}$ in the ACSVCS is a little smaller than our result. $R_{e,gc}$ in the previous study \citep{2022MNRAS.510.5725D} is consistent with that in this study.

\item NGC4461: It has a neighbor, NGC4458, so we fit two S\'ersic functions simultaneously, but it also has a large neighbor to the northeast. We masked it out.

\item VCC1185: There is no special feature in GC properties. This galaxy was studied in \citet{1996AJ....112..972D}, and their total number of GCs ($N_{GC}=7\pm9$) is smaller than our result. This discrepancy could be due to the limited spatial coverage of the previous study. The ACSVCS also studied this galaxy with a slightly smaller total number of GCs ($N_{GC} = 14\pm5.7$ than that in this study, but they are consistent within the margin of error. 

\item NGC4472: It has a neighbor, NGC4365, so we fit two S\'ersic functions simultaneously. Other small neighbors and contaminations are masked out. The GC system of this galaxy has been studied in many literature (e.g. \citealp{1998AJ....115..947L,2001AJ....121..210R,2003ApJ...591..850C}). \citet{2014ApJ...796...62H} estimated $R_{e,GC}$ with a S\'ersic profile, and their result ($R_{e,gc}=12\arcmin\pm2\arcmin$) is consistent with that in this study. The total number of GCs in the ACSVCS ($N_{GC}=7813\pm830$,\citealp{2008ApJ...681..197P}) is slightly smaller than that in this study. This discrepancy may be due to different ways of estimating the background and effect of neighbor galaxies.

\item NGC4473: There is no special feature in GC properties. Several previous studies estimated the total number of GCs. However, their results are smaller than that in this study mainly due to the small areal coverage of the previous studies \citep{2001AJ....121.2950K,2001AJ....121.2974L,2008ApJ...681..197P}.

\item NGC4474: There is no special feature in GC properties. \citet{2022MNRAS.510.5725D} estimated $R_{e,gc}$, and their result ($R_{e,gc}=0\arcmin.64\pm0\arcmin.09$) is slightly smaller than our result. $N_{gc}$ from the ACSVCS ($N_{gc}=116\pm24$) is also slightly smaller than our result. Both discrepancies could be due to small areal coverage of previous studies.

\item NGC4476: It has a huge neighbor, NGC4486, so we fit two S\'ersic functions simultaneously The $N_{GC}$ of this galaxy in the ACSVCS is slightly smaller than our result, but both are consistent within the margin of error.

\item NGC4477: It has a huge neighbor, NGC4473, so we fit two S\'ersic functions simultaneously.

\item NGC4482: There is no special feature in GC properties. $N_{GC}$ from the ACSVCS \citep{2008ApJ...681..197P} is smaller than that in this study, and it could be due to the small areal coverage of the ACSVCS.

\item NGC4478: It has a huge neighbor, NGC4486, so we fit two S\'ersic functions simultaneously. $N_{GC}$ from the ACSVCS \citep{2008ApJ...681..197P} is smaller than that in this study, and it could be due to ways to subtract the contamination from the big neighbor. 

\item NGC4479: It has a neighbor, NGC4473, so we fit two S\'ersic functions simultaneously, but it also has a large contamination to the northwest. We masked it out. $N_{GC}$ from the ACSVCS \citep{2008ApJ...681..197P} is consistent with that in this study.

\item NGC4483: There is no special feature in GC properties. \citet{2022MNRAS.510.5725D} estimated $R_{e,gc}$, and their result ($R_{e,gc}=0\arcmin.45\pm0\arcmin.02$) is smaller than our result. $N_{gc}$ from the ACSVCS ($N_{gc}=58.6\pm9.3$) is also smaller than our result. Both discrepancies could be due to small areal coverage of previous studies.

\item NGC4486: It has a lot of neighbor galaxies, but we masked them out except for NGC4406. We fit two S\'ersic functions simultaneously. Our GC radial number density profile is consistent with previous studies (Figure \ref{comp_n4486}). There are a lot of previous studies for the GC system of this galaxy, and the total number of GCs in the previous studies is about $14000-15000$ (e.g. \citealp{2006MNRAS.373..588T,2008ApJ...681..197P}), but we have a little larger $N_{GC}$. It may be due to different spatial coverage.

\item NGC4489: There is no special feature in GC properties. $N_{GC}$ from the ACSVCS \citep{2008ApJ...681..197P} is smaller than that in this study. Please check the section \ref{compacsvcs}.

\item IC3461: There is a bump on the outer edge of the GC radial number density profile. The GC system of this galaxy was studied in \citet{1996AJ....112..972D}, and their $N_{GC}$ ($=16\pm12$) is smaller than that of this study, but both results are consistent within the margin of error.

\item NGC4503: It has a neighbor, IC3470, so we fit two S\'ersic functions simultaneously.

\item IC3468: The blue and red GC populations are comparable, but blue and red peaks are close. $N_{GC}$ from the ACSVCS \citep{2008ApJ...681..197P} is smaller than that in this study. Please check the section \ref{compacsvcs}.

\item IC3470: There is no special feature in GC properties. $N_{GC}$ from the ACSVCS \citep{2008ApJ...681..197P} is smaller than that in this study. Please check the section \ref{compacsvcs}.

\item IC798: There is no special feature in GC properties. $N_{GC}$ from the ACSVCS \citep{2008ApJ...681..197P} is smaller than that in this study. Please check the section \ref{compacsvcs}.

\item NGC4515: There is no special feature in GC properties. $R_{e,gc}$ from the previous study ($R_{e,gc}=0\arcmin.40 \pm 0\arcmin.06$, \citealp{2019MNRAS.485..382C}) is slightly smaller than our result, but they are consistent within the margin of error. $N_{GC}$ from the ACSVCS \citep{2008ApJ...681..197P} is smaller than that in this study. Please check the section \ref{compacsvcs}.

\item VCC1512: GCs are rarely detected. $N_{GC}$ from the ACSVCS \citep{2008ApJ...681..197P} is smaller than that in this study. Please check the section \ref{compacsvcs}.

\item IC3501: There are three bright stars nearby, so we masked them out. $N_{GC}$ from the ACSVCS \citep{2008ApJ...681..197P} is smaller than that in this study. Please check the section \ref{compacsvcs}.

\item NGC4528: There is a bright star in the east and a dwarf galaxy in the north, so we masked them out. $N_{GC}$ from the ACSVCS \citep{2008ApJ...681..197P} is smaller than that in this study. Please check the section \ref{compacsvcs}.

\item VCC1539: There is no special feature in GC properties. $N_{GC}$ from the ACSVCS \citep{2008ApJ...681..197P} is smaller than that in this study, but they are consistent within the margin of error. Please check the section \ref{compacsvcs}.

\item IC3509: There is no special feature in GC properties. $N_{GC}$ from the ACSVCS \citep{2008ApJ...681..197P} is smaller than that in this study, but they are consistent within the margin of error. Please check the section \ref{compacsvcs}.

\item NGC4550: It has a neighbor, NGC4551, so we fit two S\'ersic functions simultaneously. The blue and red GC populations are comparable, which is similar to the results in \citet{2006ApJ...639...95P}. $N_{GC}$ from previous studies \citep{2001AJ....121.2950K,2008ApJ...681..197P} are consistent with that in this study.

\item NGC4551: It has a neighbor, NGC4550, so we fit two S\'ersic functions simultaneously. Large area at North is masked out due to a large neighbor galaxy. The blue and red GC populations are comparable, which is similar to the results in \citet{2006ApJ...639...95P}. $N_{GC}$ from the previous study \citep{2008ApJ...681..197P} is consistent with that in this study.

\item NGC4552: There are several dwarf galaxies nearby, so we masked them out. $R_{e,gc}$ from the previous study ($R_{e,gc}=2\arcmin.6 \pm 1\arcmin.0$, \citealp{2019MNRAS.485..382C}) is much smaller than our result, and $N_{GC}$ from the ACSVCS \citep{2008ApJ...681..197P} is also smaller than that in this study. It may be due to the limited spatial coverage of previous studies. Please check the section \ref{compacsvcs}.

\item VCC1661: There is no special feature in GC properties. $N_{GC}$ from the ACSVCS \citep{2008ApJ...681..197P} is smaller than that in this study. Please check the section \ref{compacsvcs}.

\item NGC4564: There is a bright star in the west and a dwarf galaxy in the south, so we masked them out. The red GC population is larger than the blue GC population, which is already shown in the ACSVCS \citep{2006ApJ...639...95P}. $R_{e,gc}$ from the previous study ($R_{e,gc}=0\arcmin.6 \pm 0\arcmin.1$, \citealp{2019MNRAS.485..382C}) is slightly smaller than our result, but they are consistent within the margin of error. $N_{GC}$ from the ACSVCS \citep{2008ApJ...681..197P} is consistent with that in this study

\item NGC4570: There is no special feature in GC properties. $R_{e,gc}$ from the previous study ($R_{e,gc}=1\arcmin.6 \pm 0\arcmin.5$, \citealp{2022MNRAS.510.5725D}) is slightly larger than our result, but they are consistent within the margin of error. $N_{GC}$ from the ACSVCS \citep{2008ApJ...681..197P} is smaller than that in this study. Please check the section \ref{compacsvcs}.

\item NGC4578: There is a bright star in the east, so we masked it out. The $N_{GC}$ of this galaxy in the ACSVCS \citep{2008ApJ...681..197P} is smaller than our result, but they are consistent within the margin of error.
 
\item NGC4596: There is a bump on the outer edge of the GC radial number density profile. 

\item VCC1826: GCs are rarely detected. The $N_{GC}$ of this galaxy in the ACSVCS \citep{2008ApJ...681..197P} is smaller than our result, but they are consistent within the margin of error.

\item VCC1833: The GC number density peak is offset from the galaxy center. The $N_{GC}$ of this galaxy in the ACSVCS \citep{2008ApJ...681..197P} is slightly smaller than our result, but they are consistent within the margin of error.

\item IC3647: The GC color distribution is determined to be bimodal by statistical testing, although the number of GCs is small and it was determined as an unimodal distribution in the ACSVCS \citep{2006ApJ...639...95P}. The $N_{GC}$ of this galaxy in the ACSVCS is slightly smaller than our result, but they are consistent within the margin of error.

\item IC3652: There is no special feature in GC properties. The $N_{GC}$ of this galaxy in the ACSVCS \citep{2008ApJ...681..197P} is smaller than our result, but they are consistent within the margin of error.

\item NGC4608: The GC spatial distribution is elongated.

\item IC3653: It has a neighbor, NGC4621, so we fit two S\'ersic functions simultaneously. The $N_{GC}$ of this galaxy in the ACSVCS \citep{2008ApJ...681..197P} is slightly smaller than our result, but they are consistent within the margin of error.

\item NGC4612: This galaxy is on the edge of the NGVS footprint. $N_{GC}$ from the ACSVCS \citep{2008ApJ...681..197P} is much smaller than that in this study. Please check the section \ref{compacsvcs}.

\item VCC1886: GCs are rarely detected. The $N_{GC}$ of this galaxy in the ACSVCS \citep{2008ApJ...681..197P} is slightly smaller than our result, but they are consistent within the margin of error.

\item UGC7854: The GC number density peak is offset from the galaxy center. The blue and red GC populations are comparable, but it was shown as an unimodal in the ACSVCS \citep{2006ApJ...639...95P}. The $N_{GC}$ of this galaxy in the ACSVCS \citep{2008ApJ...681..197P} is slightly smaller than our result, but they are consistent within the margin of error.

\item NGC4621: It has a neighbor, NGC4649, so we fit two S\'ersic functions simultaneously. $R_{e,gc}$ from the previous study ($R_{e,gc}=7\arcmin.1 \pm 1\arcmin.3$, \citealp{2019MNRAS.485..382C}) is larger than our result. $N_{GC}$ from the ACSVCS \citep{2008ApJ...681..197P} is smaller than that in this study.  These discrepancies could be due to ways to subtract the contamination from the big neighbor.  

\item NGC4638: It has a neighbor, NGC4649, so we fit two S\'ersic functions simultaneously. $N_{GC}$ from the ACSVCS \citep{2008ApJ...681..197P} is much smaller than that in this study. Please check the section \ref{compacsvcs}.

\item NGC4649: It has a neighbor, NGC4621, so we fit two S\'ersic functions simultaneously. The total number of GCs is $8875^{+508}{-419}$, which is twice the number reported in the literature ($N{GC}=4745 \pm 1099$, ACSVCS; $N_{GC}=3700 \pm 900$, \citealp{2004MNRAS.355..608F}; $N_{GC}=3600 \pm 500$, \citealp{2008ApJ...682..135L}). This discrepancy is mainly due to background estimation. The background was defined in the literature as approximately 10 arcminutes from the galaxy center, but we found that $R_{e,gc}$ is larger than 10 arcminutes ($R_{e,gc}=15\arcmin.71^{+0.41}_{-0.38}$). The excess of GCs over the background is very clear up to tens of arcminutes (Fig. \ref{fig:n4649}). However, the GC number density profile does not seem to fit well with a single S\'ersic profile. There may be an additional component starting at approximately 10 arcminutes radius. While we cannot confirm this second component, we do observe a higher number of GCs in this galaxy.

\item VCC1993: GCs are rarely detected. The GC number density peak is offset from the galaxy center. $N_{GC}$ from the ACSVCS \citep{2008ApJ...681..197P} is negative, but we have a positive number. Please check the section \ref{compacsvcs}.

\item NGC4660: It has a neighbor, NGC4649, so we fit two S\'ersic functions simultaneously. $R_{e,gc}$ from the previous study ($R_{e,gc}=0\arcmin.7 \pm 0\arcmin.1$, \citealp{2019MNRAS.485..382C}) is much smaller than our result, and $N_{GC}$ from the ACSVCS \citep{2008ApJ...681..197P} is also smaller than that in this study. It may be due to the limited spatial coverage of previous studies. Please check the section \ref{compacsvcs}.

\item IC3735: There is no special feature in GC properties. The $N_{GC}$ from the ACSVCS \citep{2008ApJ...681..197P} is consistent with that in this study.

\item IC3773: There is no special feature in GC properties. The $N_{GC}$ from the ACSVCS \citep{2008ApJ...681..197P} is consistent with that in this study.

\item IC3779: There is no special feature in GC properties. The $N_{GC}$ of this galaxy in the ACSVCS \citep{2008ApJ...681..197P} is slightly smaller than our result, but they are consistent within the margin of error.

\item NGC4694: The red GC population is larger than the blue GC population.

\item NGC4710: This galaxy is on the edge of the NGVS footprint. \citet{2010ApJ...721..893M} detected 63 GC candidates using HST/ACS image.

\item NGC4733: The GC number density peak is a little offset from the galaxy center.

\item NGC4754: There is a neighbor galaxy in the east, so we masked it out. The blue and red GC populations are comparable. The $N_{GC}$ from previous studies ($N_{GC}=115\pm15$, \citealp{2012AJ....144..164H}; $N_{GC}=103\pm17$, \citealp{2008ApJ...681..197P}) are slightly smaller than our result, but they are consistent within the margin of error. The $R_{e,gc}$ from previous studies ($R_{e,gc}=2\arcmin.6\pm0\arcmin.9$, \citealp{2012AJ....144..164H,2019MNRAS.485..382C}) is larger than that in this study.

\item NGC4762: There is a neighbor galaxy in the east, so we masked it out. This galaxy is on the edge of the NGVS footprint. The $N_{GC}$ from previous studies ($N_{GC}=270\pm30$, \citealp{2012AJ....144..164H}; $N_{GC}=211\pm34$, \citealp{2008ApJ...681..197P}) are slightly smaller than our result, but they are consistent within the margin of error. The $R_{e,gc}$ from previous studies ($R_{e,gc}=1\arcmin.4\pm0\arcmin.4$, \citealp{2012AJ....144..164H,2019MNRAS.485..382C}) is consistent with that in this study.

\item NGC5839: It has a neighbor, NGC5846, so we fit two S\'ersic functions simultaneously.

\item NGC5846: It has a neighbor, NGC5839, so we fit two S\'ersic functions simultaneously. The blue and red GC populations are comparable. \citet{1997AJ....113..887F} studied the GC system of this galaxy based on the HST/WFPC2 observation, and they provided $N_{GC}=4670$ based on the extrapolation. This value is much larger than our result, and it could be due to the limited spatial coverage and extrapolation of the previous study. Interestingly, \citet{1997AJ....113..887F} estimated $S_N$ to be $4.1\pm1.1$ which is much smaller than our result even with larger $N_{GC}$. This discrepancy is mainly due to the different magnitudes of the host galaxy. 

\item NGC5866: GCs are more centrally concentrated than other galaxies. There are several previous studies on the GC system of this galaxy \citep{2007ApJ...668..209C,2012AJ....144..164H}, and the $N_{GC}$ of the previous studies ($N_{GC}\sim300$, \citealp{2007ApJ...668..209C}; $N_{GC}=340\pm80$, \citealp{2012AJ....144..164H}) are consistent with our result. $R_{e,gc}$ was also estimated in previous studies ($R_{e,gc}=3\arcmin.1\pm0\arcmin.7$, \citealp{2012AJ....144..164H,2019MNRAS.488.4504C}), but the value from the literature is much larger than that in this study.

\item PGC058114: The GC number density peak is offset from the galaxy center.

\item NGC6548: It contains only one GC within $2R_{e,gc}$.

\item NGC7280: It contains no GC within $2R_{e,gc}$.

\item NGC7332: There is no special feature in GC properties. $N_{GC}$ from previous studies ($N_{GC}=190\pm30$, \citealp{2001MNRAS.325.1431F}; $N_{GC}=175\pm15$, \citealp{2012AJ....144..103Y}) are larger than that in this study, but they are consistent within the margin of error. $R_{e,gc}$ from the previous study \citep{2018MNRAS.477.3869H} is consistent with that of this study. 

\item NGC7457: There is no special feature in GC properties. $N_{GC}$ from the previous study ($N_{GC}=210\pm30$, \citealp{2011ApJ...738..113H}) is larger than that in this study, but they are consistent within the margin of error.

\item NGC7454: GCs are more centrally concentrated than other galaxies.

\end{itemize}
\section{Diagnostic plots for all samples}
\figsetstart
\figsetnum{10}


\figsetgrpstart
\figsetgrpnum{10.1}
\figsetgrptitle{NGC1023}
\figsetplot{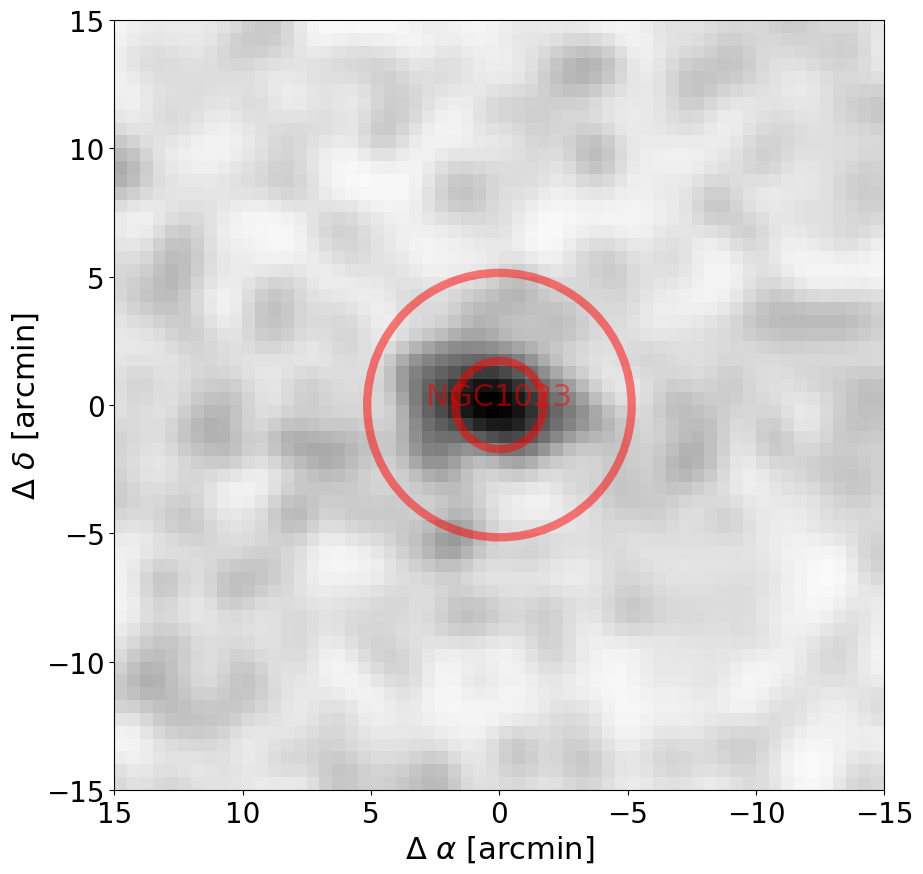}
\figsetplot{corner_n1023.png}
\figsetplot{agcprof_n1023.png}
\figsetplot{colgc_n1023.png}
\figsetgrpnote{NGC1023. See Figure \ref{fig:n0821} for details.}
\figsetgrpend

\figsetend





\begin{figure*}
\gridline{\fig{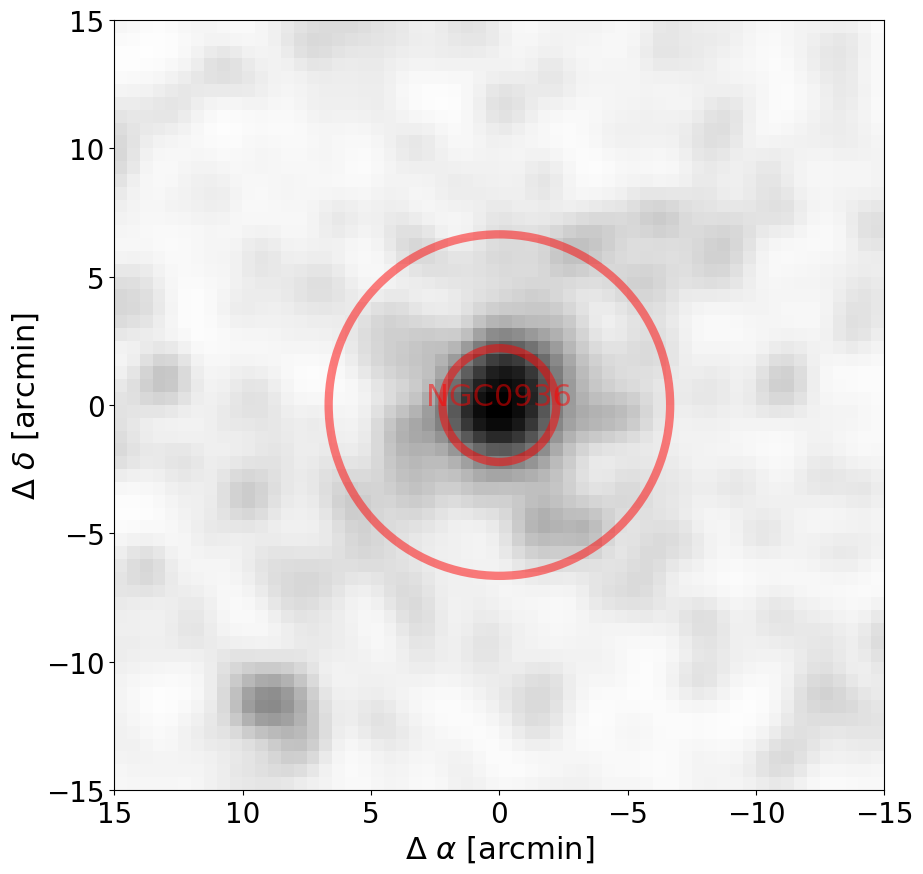}{0.48\textwidth}{(a)}
          \fig{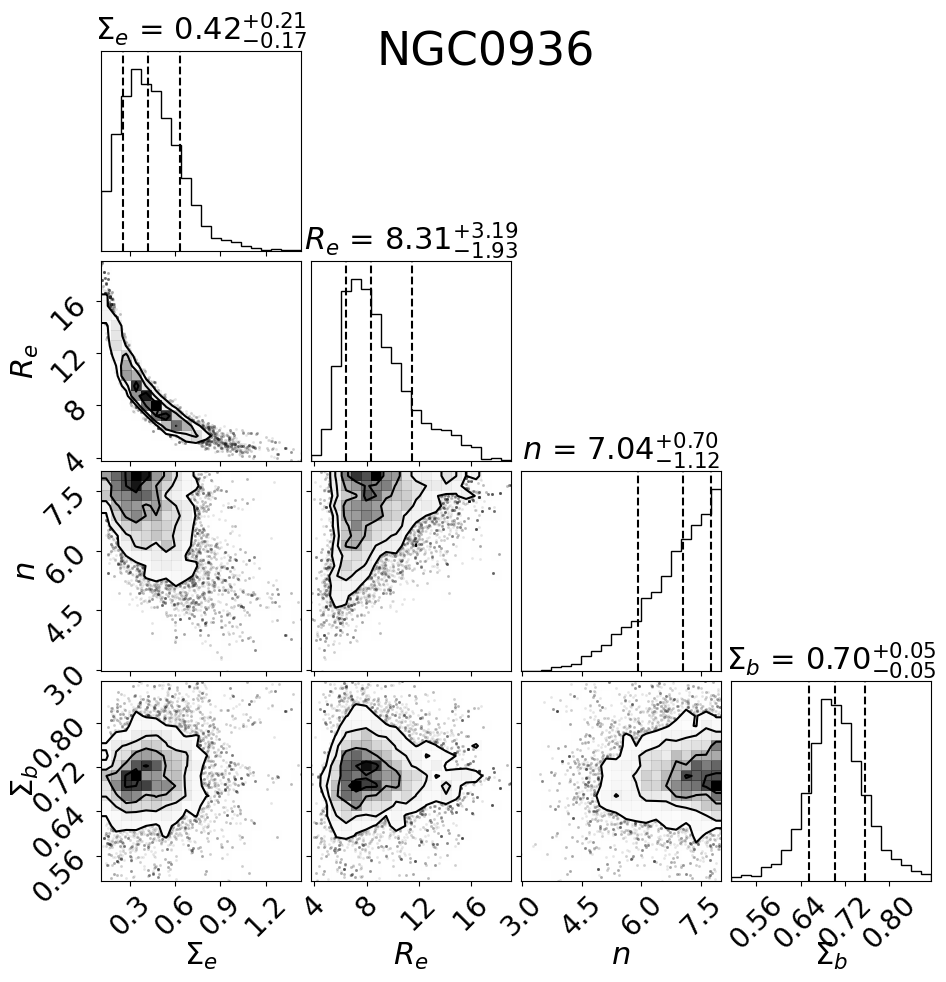}{0.48\textwidth}{(b)}
          }
\gridline{\fig{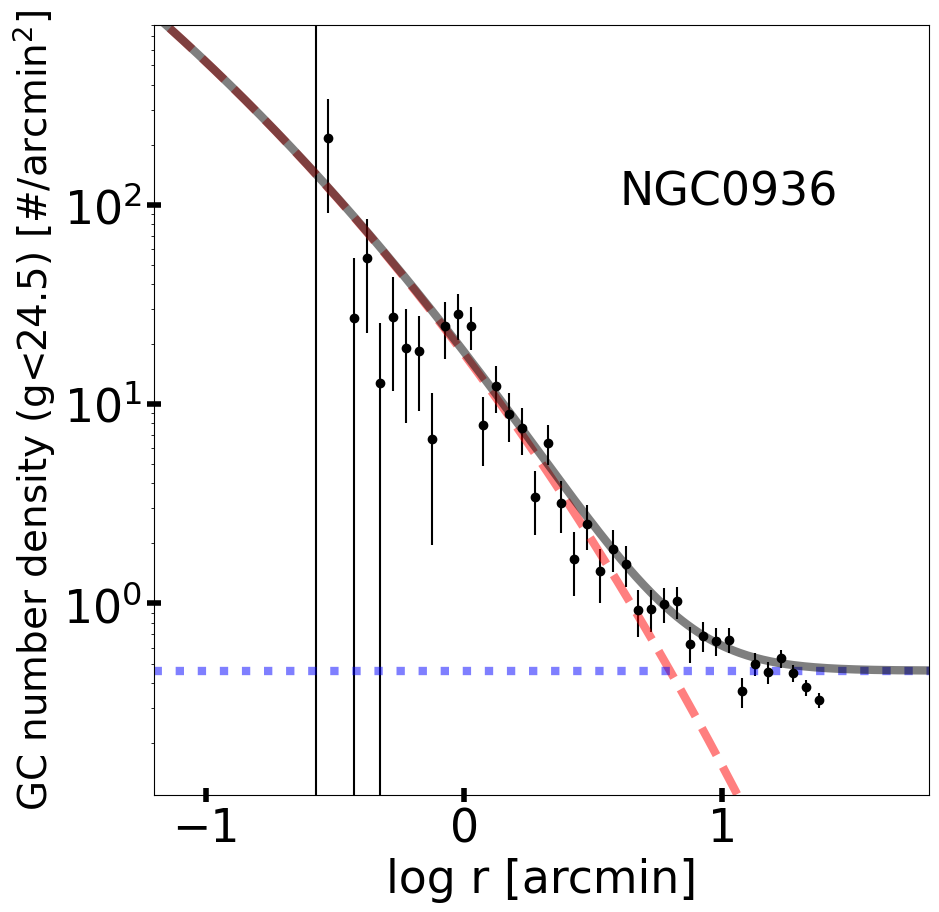}{0.48\textwidth}{(c)}
          \fig{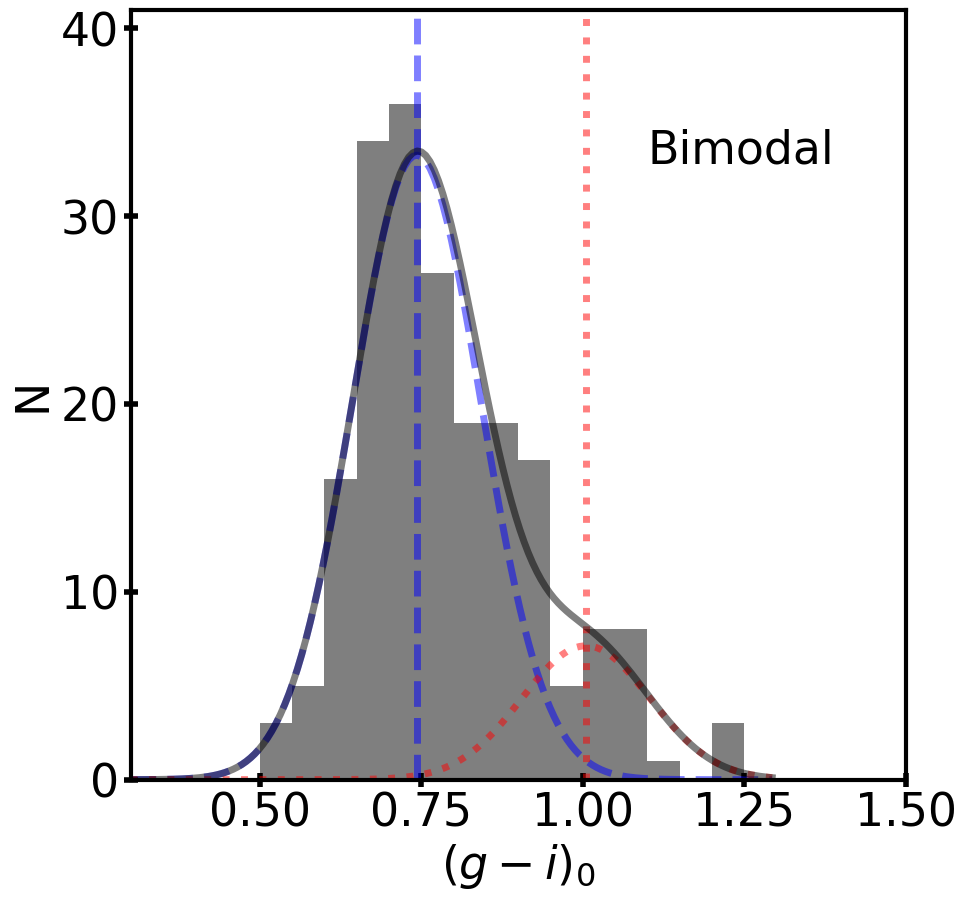}{0.48\textwidth}{(d)}
          }
\caption{NGC936. See Figure \ref{fig:n0821} for details. The complete figure set (115 images) is available in the online journal
\label{fig:n0936}}
\end{figure*}

\bibliography{ms}{}
\bibliographystyle{aasjournal}



\end{document}